\documentclass[aps, prd, twocolumn, showpacs, superscriptaddress, groupedaddress]{revtex4} 
\usepackage{graphicx}	
\usepackage{amssymb}
\usepackage{dcolumn}
\usepackage{color}
\usepackage{subfigure, rotating, bm, array}
\usepackage[pagebackref=false, colorlinks=true]{hyperref}
\hypersetup{
linkcolor=blue,     
citecolor=blue,     
urlcolor=blue}      
\begin{document}
\title{Timelike Geodesics in Naked Singularity and Black Hole Spacetimes II}
\author{Ashok B. Joshi}
\email{gen.rel.joshi@gmail.com}
\affiliation{International Center for Cosmology, Charusat University, Anand 388421, Gujarat, India}
\author{Parth Bambhaniya}
\email{grcollapse@gmail.com}
\affiliation{International Center for Cosmology, Charusat University, Anand 388421, Gujarat, India}
\author{Dipanjan Dey}
\email{dipanjandey.adm@charusat.edu.in}
\affiliation{International Center for Cosmology, Charusat University, Anand 388421, Gujarat, India}
\author{Pankaj S. Joshi}
\email{psjprovost@charusat.ac.in}
\affiliation{International Center for Cosmology, Charusat University, Anand 388421, Gujarat, India}

\date{\today}

\begin{abstract}
We derive here the orbit equations of particles in naked singularity spacetimes, namely the Bertrand (BST) and Janis-Newman-Winicour (JNW) geometries, and for the Schwarzschild black hole. We plot the orbit equations and find the Perihelion precession of the orbits of particles in the BST and JNW spacetimes and compare these with the Schwarzschild black hole spacetime. We find 
and discuss different distinguishing properties in the effective potentials and orbits of particle in BST, JNW and Schwarzschild spacetimes, and the particle trajectories are shown for the matching of BST with an external Schwarzschild spacetime. We show 
that the nature of perihelion precession of orbits in BST and Schwarzschild spacetimes are similar, while in the JNW case the nature of perihelion precession of orbits is opposite to that of
the Schwarzschild and BST spacetimes. Other interesting and important features of these orbits are pointed out.
\\
\\
$\boldsymbol{key words}$ : Naked singularity, Black hole, Perihelion precession, Galaxy, Universe.

\end{abstract}
\maketitle

\section{Introduction}
In General theory of Relativity (GR), many predictions have been tested by observational evidence, such as the precession of  perihelion of orbits of mercury \cite{precession}, the bending of light near the sun \cite{dyson}, gravitational red shift \cite{rebka}, and more recently the  discovery of gravitational waves \cite{ligo}. In such a context, the gravitational collapse of a massive star is the most interesting and fascinating phenomenon in the universe. What is the final fate of the massive star? Dynamical evolution of massive stars within the framework of Einstein theory of gravity predicts that a spacetime singularity must arise as a collapse end state, and ultra-strong gravity regions form 
\cite{Joshi:2004sf, OppenheimerSnyder39, Datt,Joshi:2011zm,JMN}. 
Such very strong gravity regions also exist  at the center of a galaxy where so much matter is compacted in a very small region. 

As for predicting the final fate of massive stars in Einstein gravity, 
in 1939, Oppenheimer and Snyder, and Dutt, gave for the first time 
a dynamic collapse model (OSD) \cite{OppenheimerSnyder39},\cite{Datt},
where the star finally terminates into a black hole as the end state of gravitational collapse. For simplicity they took non-realistic assumptions such as the density distribution 
being entirely homogeneous, the gas pressure being totally neglected within the star, and such others.
In this case, the event horizon and apparent horizon formed before the formation of central strong singularity which is hidden within 
the black hole. However, if we consider a more realistic inhomogeneous collapse, with density higher at center and decreasing slowly as we move away, the central strong singularity forms before the formation of event horizon and the apparent horizon, so the singularity is at least locally visible, also called a   naked singularity   \cite{Joshi:1993zg,Singh:1994tb,Mena:1999bz}. 
More generally, many recent works have shown that small inhomogeneity or pressures inside the collapsing matter cloud 
allow for strong curvature central singularity in Einstein gravity, 
which is locally or globally visible. 

The black hole as well as naked singularities, if they occur in nature, would be physically and causally very different entities, and may have quite different and distinct astrophysical signatures. To explore such possibilities for some distinguishable observational signatures, 
we consider here two specific naked singularity spacetimes as test cases, namely the 
Bertrand and JNW geometries. In \cite{Dey:2013yga},\cite{Dey+15} astrophysical importance of Bertrand spacetimes was 
discussed. The gravitational lensing and shadow due to JNW naked singularity is discussed in \cite{Gyulchev:2019tvk}, which closely resembles with the shadow of a Schwarzschild black hole. Gravitational lensing and accretion disk properties for these objects are  investigated in \cite{Joshi:2013dva}, \cite{Kovacs:2010xm}. 
Also, recently the black hole shadow was discovered 
\cite{M87}. It was shown that similar shadow can be created by the  JMN naked singularity for some cases \cite{Shaikh:2018lcc}. In \cite{shaikh1, shaikh2, shaikh3, sunny}, gravitational lensing and shadow due to the ultra compact objects, wormholes and superspinars, is investigated.  These are all interesting and useful theoretical predictions
in the context of the recent observation of shadow of the M87 galactic center \cite{Akiyama:2019fyp}. 

It follows that the theoretical predictions of possible observational signatures of black hole and naked singularity spacetimes, and their differences and similarities are worth exploring. What we need is 
to understand carefully these compact region geometries and their  
properties. In such a perspective, understanding the perihelion precession of a particle in such spacetimes could be a challenging
and useful issue in GR, which is usually treated in terms of the timelike geodesics that the particles move along in a given spacetime. 

As we know, Einstein  investigated the timelike geodesics of a particle in Schwarzschild spacetime \cite{schprecession}, and he found the well-known perihelion precession formula, $$\Delta\phi = \frac{6\pi G M}{c^2~a(1-e^2)}\,\, ,$$  where $G$ is the gravitational constant, $M$ is the total mass of the central object, $a$ is a semi-major axis and $e$ is the eccentricity of orbit. To calculate the perihelion precession, approximation method for weak gravitational field was suggested by Kerner et.al.\cite{kerner}. Higher order geodesic deviations and orbital precession in Kerr-Newman spacetime was investigated in \cite{heydari}.
 Charged particles moving along circular orbits around the central body
were examined to distinguish the black hole and naked singularity \cite{remo1}. Also, circular stable and unstable orbits around configurations describing either black holes or naked singularities 
were considered in \cite{remo2}. In a recent work we also presented 
a study involving timelike geodesics in so called JMN naked singularity spacetimes, comparing it with black hole case  
\cite{Parth}. 

From an observational perspective, it may be worth noting that recently the UCLA galactic center group demonstrated that short-period stars (e.g. S0-102 and S0-2) orbiting around the super-massive black hole in our galactic center can be successfully  used to probe the gravitation theory in a strong field regime \cite{UCLA}. Over past 17 years, the W.M. Keck observatory was used to image the galactic center at the highest angular resolution possible today. They have detected S0-102, a star orbiting our galaxy’s super-massive black hole, with a period of just 11.5 \cite{shortest} years. 
Also SINFONI gave the galactic center data, and updates on monitoring stellar orbits in the galactic center of our milky way
are given \cite{UCLA,shortest,center1,Eisenhauer:2005cv}.

This observational information can help us reveal the nature of the central object SGR-A* of our Milky way galaxy, which is considered to be a super massive black hole with mass of about 
$10^6 M_{\odot}$. From such a context, it is important to do a comparative study of the nature of the timelike geodesics and perihelion precession in different singular spacetimes.
Since the BST and JNW spacetimes are static, non-vacuum solutions of Einstein equations, timelike orbits in these spacetimes should be distinguishable from the timelike orbits in the vacuum black hole spacetimes, namely the Schwarzschild geometry, and this difference is reported in the present work.  
So, we discuss here the timelike geodesics in the BST and JNW spacetimes, both of which have a central naked singularity. 

The plan of the paper is as follows. In section (\ref{BSTorbit}), we discuss the timelike geodesics in BST and Schwarzschild spacetimes in detail. We match the interior BST spacetime with exterior Schwarzschild spacetime on a hypersurface $\Sigma$, at a boundary $r=R_b$. We then compare the perihelion precession of a particle in these two spacetimes. In section (\ref{JNWorbit}), we discuss the  timelike geodesics in JNW spacetime and compare the same with Schwarzschild spacetime. As we know, the JNW spacetime is asymptotically flat, so one need not match JNW with Schwarzschild spacetime. JNW spacetime has a scalar field effect,
and  we discuss the how this scalar field changes the nature of orbits. In section (\ref{approximation}), we derive an approximation solution of orbit equations of BST, JNW, and Schwarzschild spacetimes and discuss the different distinguishable properties of particles orbits in those spacetimes. In section (\ref{Result}), we discuss the results obtained here. Finally, in section (\ref{result}), we discuss conclusion and possible future work. Throughout the paper we take $G=c=1$.

\section{Timelike geodesics in Schwarzschild and BST  spacetimes }\label{BSTorbit}
General expression of metric for a spherically symmetric, static spacetime can be written as,
\begin{equation}
    ds^2 = - g_{tt}(r)dt^2 + g_{rr}(r)dr^2 + r^2(d\theta^2 + \sin^2\theta d\phi^2)\,\,,
    \label{static}
\end{equation}
where $g_{tt}(r)$,~$g_{rr}(r)$ are the metric components of the spherical, static spacetime. The spherically symmetric, static, vacuum solution of the Einstein equations can be uniquely represented by a Schwarzschild spacetime \cite{Hartle}. 
One can write down the Schwarzschild spacetime in the following form, 
\begin{equation}
ds_{SCH}^2 = -\left(1-\frac{M_0R_b}{r}\right)dt^2 + \frac{dr^2}{\left(1-\frac{M_0R_b}{r}\right)} +r^2d\Omega^2\,\, , 
\label{SCHmetric}
\end{equation}
where $M_0, R_b$ are two constant parameters of this spacetime, and $d\Omega^2=d\theta^2+\sin^2\theta d\phi^2$ is the line element on sphere. 

The above form of Schwarzschild spacetime can be used also when there exist a spacetime structure which has internal non-vacuum spacetime and external static vacuum, which is Schwarzschild spacetime. Using the above form of Schwarzschild spacetime, the Schwarzschild radius ($R_s$) can be written as $R_s=M_0R_b$, therefore, total Schwarzschild mass $M_{TOT}=\frac{M_0R_b}{2}$. The radial distance $R_b$ is the matching radius, where a static, spherically symmetric, non-vacuum spacetime can be glued with an external Schwarzschild geometry. For this type of spacetime structure, we need $R_b>R_s$, and therefore, $0<M_0<1$. When a collapsing matter cloud virializes before the formation of a black hole, then the above mentioned spacetime structure can be formed. There are many literature where the final state of collapsing matter cloud is discussed elaborately \cite{Jhingan:1996jb, Joshi:2015mwa, Satin:2014ima, Malafarina:2010xs, Joshi:2013dva, Bhattacharya:2017chr, Dey:2019fja, Dey:2019wwh}. In \cite{Banik:2016qvf}, it is shown that Bertrand spacetimes (BST) can be formed as an equilibrium state of gravitational collapse. The line element of BST can be written as,   
\begin{eqnarray} 
 ds^2_{BST} &=& - \left(\frac{2\beta^2}{1+\frac{Rb}{r}}\right)dt^2 + \frac{dr^2}{\beta^2} + r^2d\Omega^2\,\, ,
 \label{BSTmetric}
\end{eqnarray}
where, $\beta$ and $R_b$ are the constant parameters of this spacetime. Perlick \citep{perlick} first discovered this spacetime which admits closed, stable, circular orbits passing through each point of the spacetime. This spacetime has a central strong curvature singularity which is not covered by an event horizon. 

From the above metric (eq.~(\ref{BSTmetric})), it can be seen that the BST is not asymptotically flat. Therefore, to describe an internal spacetime of a compact object by BST spacetime, we need to match this spacetime with an external Schwarzschild spacetime at a timelike hypersurface. In general relativity, maintaining two junction conditions \cite{Eric}, one can smoothly glue two spacetimes at a spacelike or timelike hypersurface. The first condition is that the induced metric from the two sides (internal and external) should be identical on a matching hypersurface, whereas the second condition states that the extrinsic curvature from the two sides should be identical on that matching hypersurface. Extrinsic curvature of a hypersurface, which is embedded in higher dimensional manifold, can be expressed in terms of the covariant derivative of normal vectors on that hypersurface: $K_{ab}=e^{\alpha}_ae^{\beta}_{b}\nabla_{\alpha}\eta_{\beta}$, where $e^{\alpha}_a$ is the tangent vector on the hypersurface and $\eta^{\beta}$ is the normal to that hypersurface. Now, if we want to glue BST with Schwarzschild spacetime (eq.~(\ref{BSTmetric}),(\ref{SCHmetric})) smoothly at the timelike hypersurface, $\Psi=r-R_b=0$, we need to satisfy the following conditions \cite{Dey:2013yga},
\begin{equation}
\beta^2=1-M_0\,\, ,\,\,M_0=\frac13\,\, ,
\label{matchcon}
\end{equation}  
where the first condition comes from induced metric matching and the second condition comes from the extrinsic curvature matching. One can verify that with the above junction conditions, the radial pressure of BST becomes zero at the matching hypersurface $\Psi$.

\subsection{Analytic solution of orbit equation in Schwarzschild and BST spacetimes }
The angular part of the line elements, in eq.~(\ref{BSTmetric}),(\ref{SCHmetric}), shows spherical symmetry of the Schwarzschild and BST spacetimes. Therefore, the angular momentum of a freely falling particle is conserved. The conservation of energy of the particle implies temporal symmetries in static BST and Schwarzschild spacetimes. The conserved angular momentum ($h$) and energy ($\gamma$) of a particle per unit mass is given by, 
\begin{eqnarray}
    h_{SCH} = r^2\frac{d\phi}{d\tau}\,\, ,\,\,\,
   \gamma_{SCH} =  \left(1-\frac{M_0R_b}{r}\right)\frac{dt}{d\tau}\,\, ,
   \label{consch}
\end{eqnarray}
\begin{eqnarray}
 h_{BST} = r^2\frac{d\phi}{d\tau}\,\, ,\,\,\,
   \gamma_{BST} =  \left(\frac{2\beta^2}{1+\frac{Rb}{r}}\right)\frac{dt}{d\tau}\,\, ,
   \label{conBST}
\end{eqnarray}
where $h_{SCH}$, $\gamma_{SCH}$ and $h_{BST}$, $\gamma_{BST}$ are the conserved angular momentum and energy per unit rest mass of a freely falling particle in Schwarzschild and BST spacetimes respectively. Here, $\tau$ is the proper time of the particle. 

We know that the normalization of four-velocity for timelike geodesics is,  $v_{\mu}v^{\mu}=-1$. From the normalization of four-velocity of a freely falling massive particle, we can write the following effective potentials for Schwarzschild and BST spacetimes,
 \begin{equation}
    (V_{eff})_{SCH}= \frac{1}{2}\left[\left(1 -\frac{M_0R_b}{r}\right)\left(1 + \frac{h_{SCH}^2}{r^2}\right) - 1\right]\,\, 
    \label{veffsch}
\end{equation}
\begin{equation}
    (V_{eff})_{BST} = \frac{1}{2}\left[\left(\frac{2\beta^2}{1+\frac{R_b}{r}}\right)\left(1 + \frac{h_{BST}^2}{r^2}\right) - 1\right]\,\, ,
    \label{veffBST}
\end{equation}
where we consider $\theta=\pi/2$ to confine the orbits of a particle on a plane. The effective potential plays a crucial role on the shape and dynamics of the trajectories of particles. The total energy ($E$) of the freely falling massive particle can be written as,
\begin{equation}
    E=\frac{g_{rr}(r)g_{tt}(r)}{2}\left(\frac{dr}{d\tau}\right)^2+V_{eff}(r)\,\, 
    \label{totalE}
\end{equation}
where $E=\frac{\gamma^2-1}{2}$. For stable circular trajectories of massive particles, one needs, $V_{eff}(r_c)=E$, $V_{eff}^{\prime}(r_c)=0$ and $V_{eff}^{\prime\prime}(r_c)>0$, where $r_c$ is the radius of the stable circular orbit. Using $V_{eff}(r)=E$, $V_{eff}^{\prime}(r)=0$, one can write down the expressions of $h$ and $\gamma$ for a circular timelike geodesic at a given radius. With $V_{eff}(r)=E$, $V_{eff}^{\prime}(r)=0$,  a massive particle can have both stable and unstable circular orbits. For stable circular orbits we need another constraint, namely, $V_{eff}^{\prime\prime}(r)>0$. Using $V_{eff}(r)=E$, we can get perihelion ($r_{min}$: the radius of minimum approach), and aphelion ($r_{max}$: maximum approach), which are points on the bound non-circular orbits of a particle. Therefore, we can define bound orbits of the freely falling massive particles in the following mathematical way,
\begin{eqnarray}
   V_{eff}(r_{min})=V_{eff}(r_{max})=E\, , \,\, \nonumber\\
   E-V_{eff}(r)>0\, ,\,\,\,\forall r\in (r_{min},r_{max}).
   \label{bound}
\end{eqnarray}

From the expression of effective potential in (\ref{veffsch}), we can derive the expression of $h_{SCH}$ and $\gamma_{SCH}$ for circular timelike geodesics,
\begin{eqnarray}
   \gamma_{SCH}^2=\frac{2\left(r-M_0R_b\right)^2}{r\left(2r-3M_0R_b\right)}\,,\, h_{SCH}^2=\frac{M_0R_br^2}{\left(2r-3M_0R_b\right)}\,\, .\label{hesch}
\end{eqnarray}
Using the expressions of conserved quantities for circular geodesics, one can show that no circular orbit is possible in the range: $0\leq r\leq \frac{3M_0R_b}{2}$. One can write this range in terms of the total mass ($M_{TOT}$) as, $0\leq r\leq 3M_{TOT}$, where $M_{TOT}=\frac{M_0R_b}{2}$. However, for stable circular orbit, we need to satisfy $ (V_{eff}^{\prime\prime})_{SCH}>0$ along with above two conditions in eq.~(\ref{hesch}). In terms of the total mass ($M_{TOT}$), the expression of $ (V_{eff}^{\prime\prime})_{SCH}$ can be written as,
\begin{eqnarray}
   (V_{eff}^{\prime\prime})_{SCH}&=&\frac{2M_{TOT}}{r}\left(\frac{6M_{TOT}-r}{3M_{TOT}-r}\right)\,\, .
   \label{isco}
\end{eqnarray}
Therefore, a massive particle can have stable circular orbit for $r\geq6M_{TOT}$ when the two conditions in eq.~(\ref{hesch})are satisfied. No stable circular orbit is possible below $r=6M_{TOT},$ and this minimum radius for stable circular orbit is known as Innermost Stable Circular Orbit (ISCO). Similarly, from the effective potential of BST spacetime (\ref{veffBST}), we can write down the following conditions for stable, circular geodesics,
\begin{eqnarray}
   \gamma^2_{BST}&=&\frac{4\beta^2r}{2r+R_b}\, , \, \\
   h^2_{BST}&=&\frac{R_br^2}{2r+R_b}\,\, , \\
   (V_{eff}^{\prime\prime})_{BST}&=&\frac{2\beta^2R_b}{2r^3+3r^2R_b+rR_b^2}>0\,\, .
\end{eqnarray}
From the above three equations, it can be understood that for $M_0<1$, unlike Schwarzschild spacetime, BST spacetime has stable circular orbits of any radius. 
For bound orbits in this spacetime, the conditions which are stated in eq.~(\ref{bound}), should be fulfilled. Now, with some given conserved value of $h$ and $E$, one can describe the shape of an orbit of a massive particle, by showing how radial coordinate $r$ changes with azimuthal coordinate $\phi$. Using eq.~(\ref{totalE}) we get,
\begin{equation}
    \frac{d\phi}{dr}=\frac{h}{r^2}\frac{\sqrt{g_{rr}(r)g_{tt}(r)}}{\sqrt{2(E-V_{eff}(r))}}\,\, .
    \label{orbitgen}
\end{equation}
Now, using the above equation one can define the following second order differential equations for Schwarzschild and BST spacetime respectively, 
\begin{equation}
   \frac{d^2u}{d\phi^2} + u - \frac{3M_0R_b}{2}u^2 - \frac{M_0R_b}{2h^2} = 0\,\, .
   \label{orbiteqsch}
\end{equation}
\begin{equation}
     \frac{d^2u}{d\phi^2} + \beta^2 u - \frac{\gamma^2 R_b}{4h^2} = 0\,\, .\label{orbiteqBST}\\
\end{equation}
where $u=\frac1r$. From the above orbit equations, one can  get the information about the shape of orbits in Schwarzschild and BST spacetimes and we can compare them.
\begin{figure*}
\subfigure[$M_0 =0.05$,$h=0.1$, $E=-0.03$]
{\includegraphics[width=70mm]{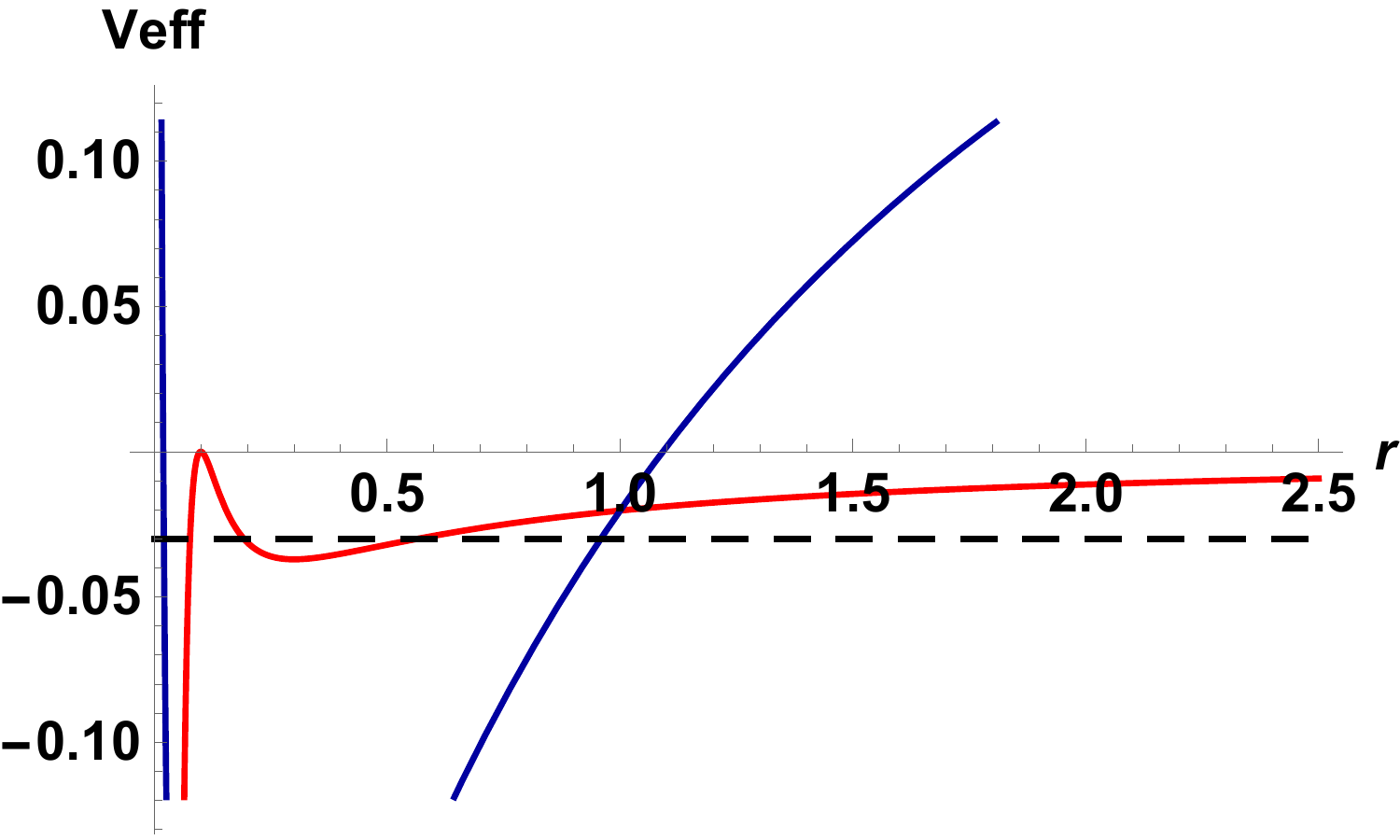}\label{veffBSTM05}}
 \subfigure[orbit in Schwarzschild spacetime]
 {\includegraphics[width=65mm]{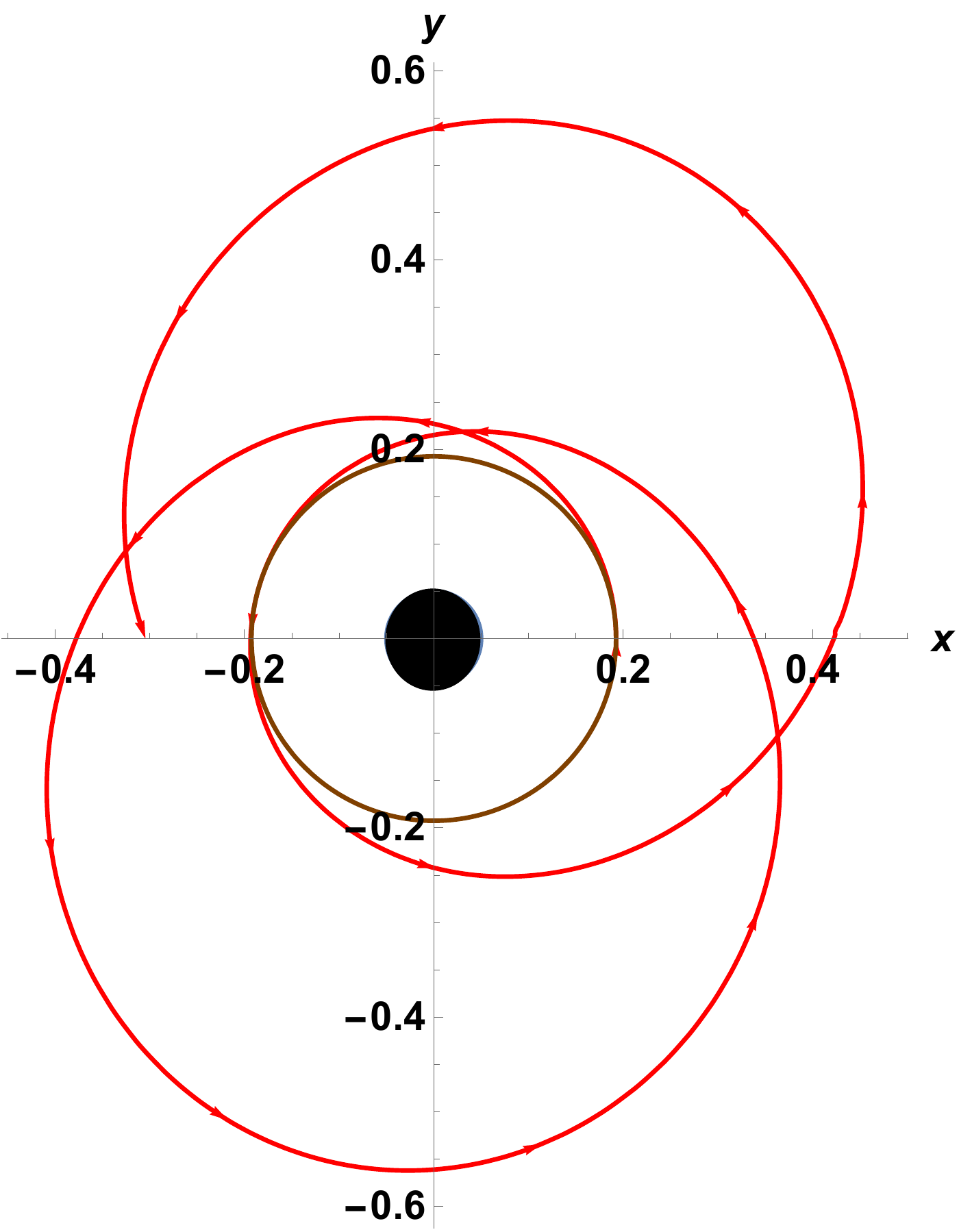}\label{orbitschM05}}
 \subfigure[orbit in BST spacetime]
 {\includegraphics[width=85mm]{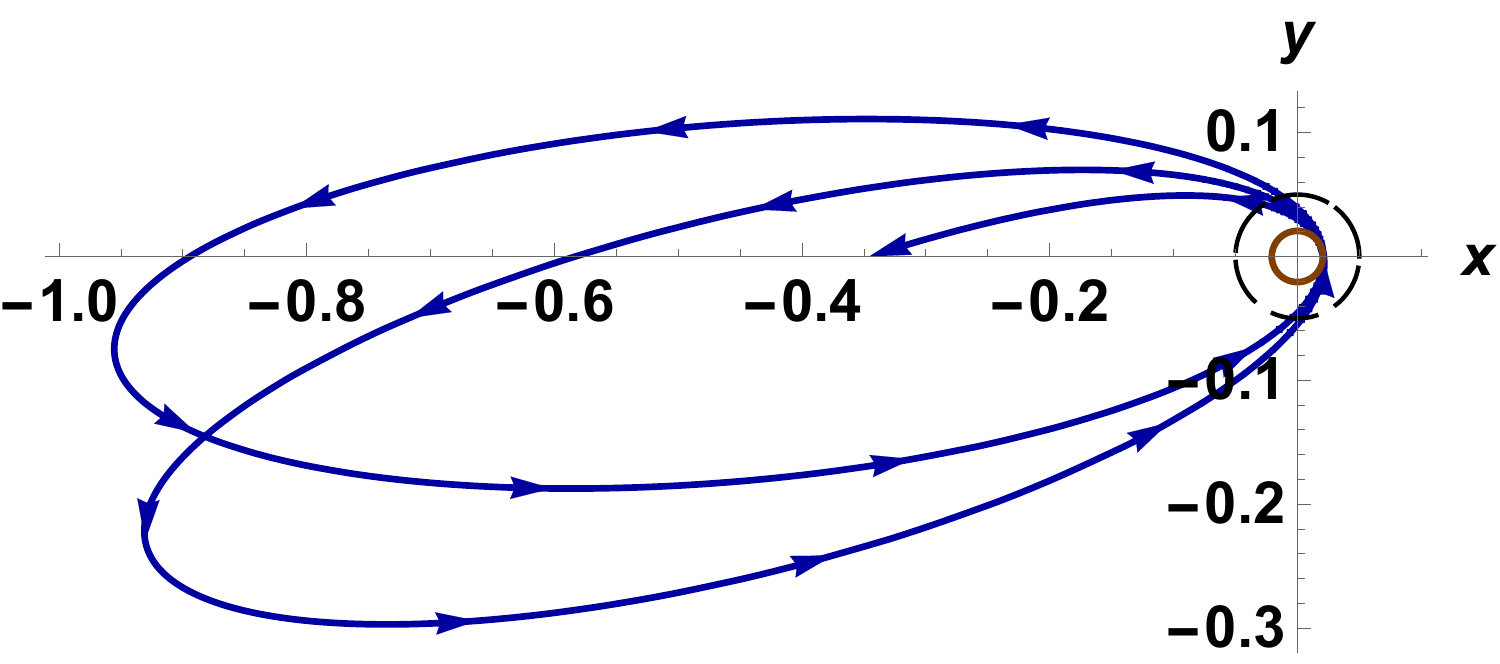}\label{orbitBSTM05}}
 \subfigure[$M_0 =0.1$,$h=0.3$, $E=-0.01$]
 {\includegraphics[width=70mm]{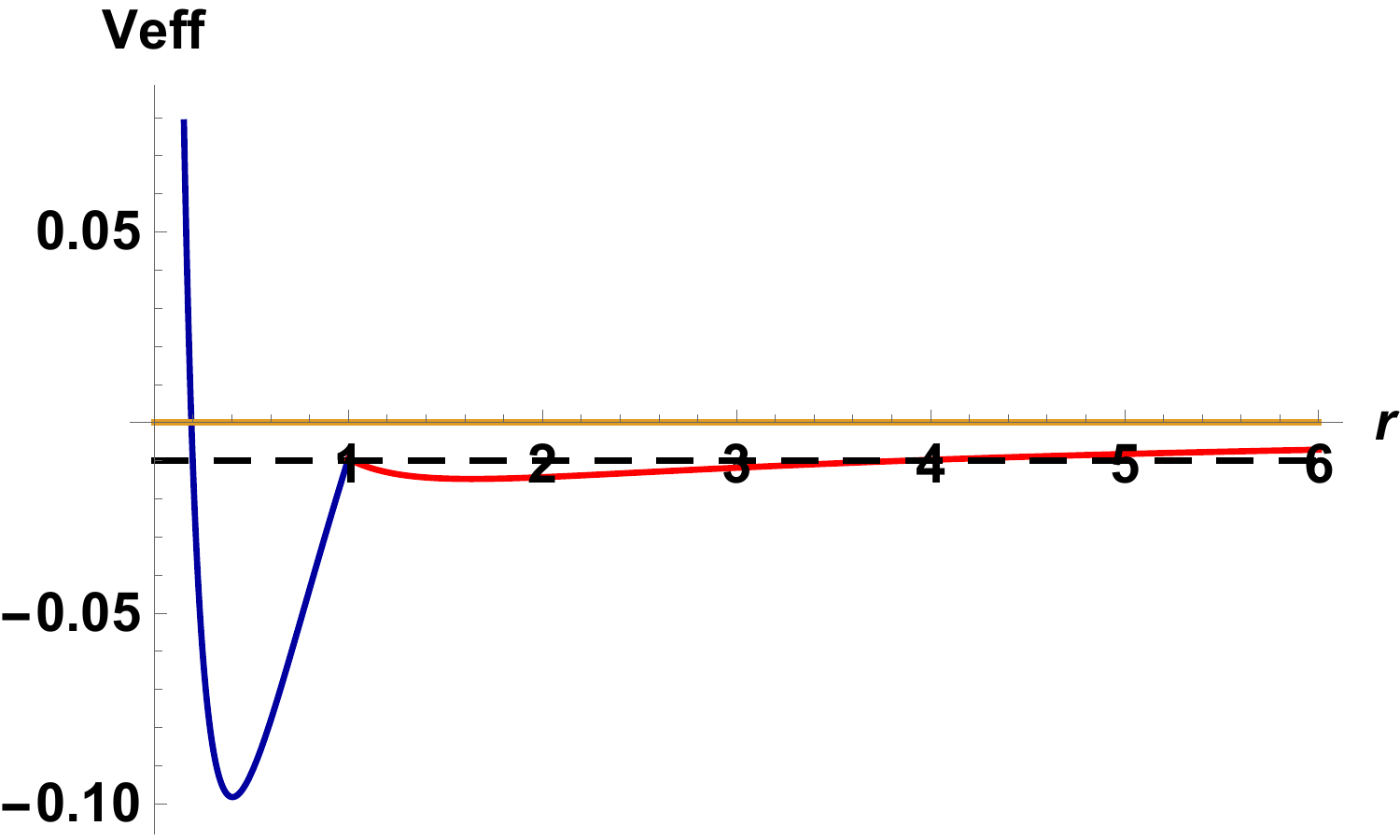}\label{veffBSTM01}}
 \subfigure[orbit in Schwarzschild spacetime]
 {\includegraphics[width=65mm]{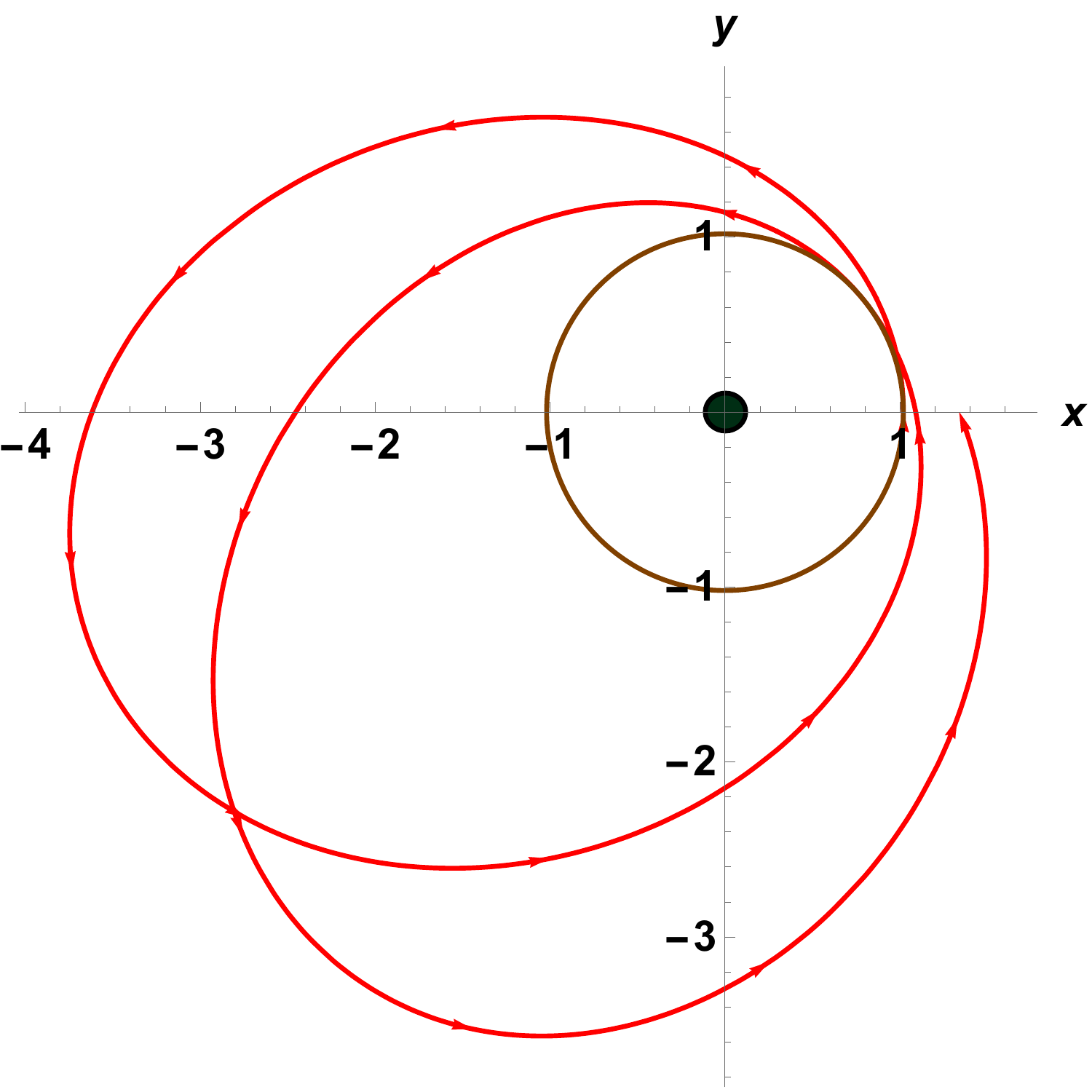}\label{orbitschM01}}
\subfigure[orbit in BST spacetime]
{\includegraphics[width=65mm]{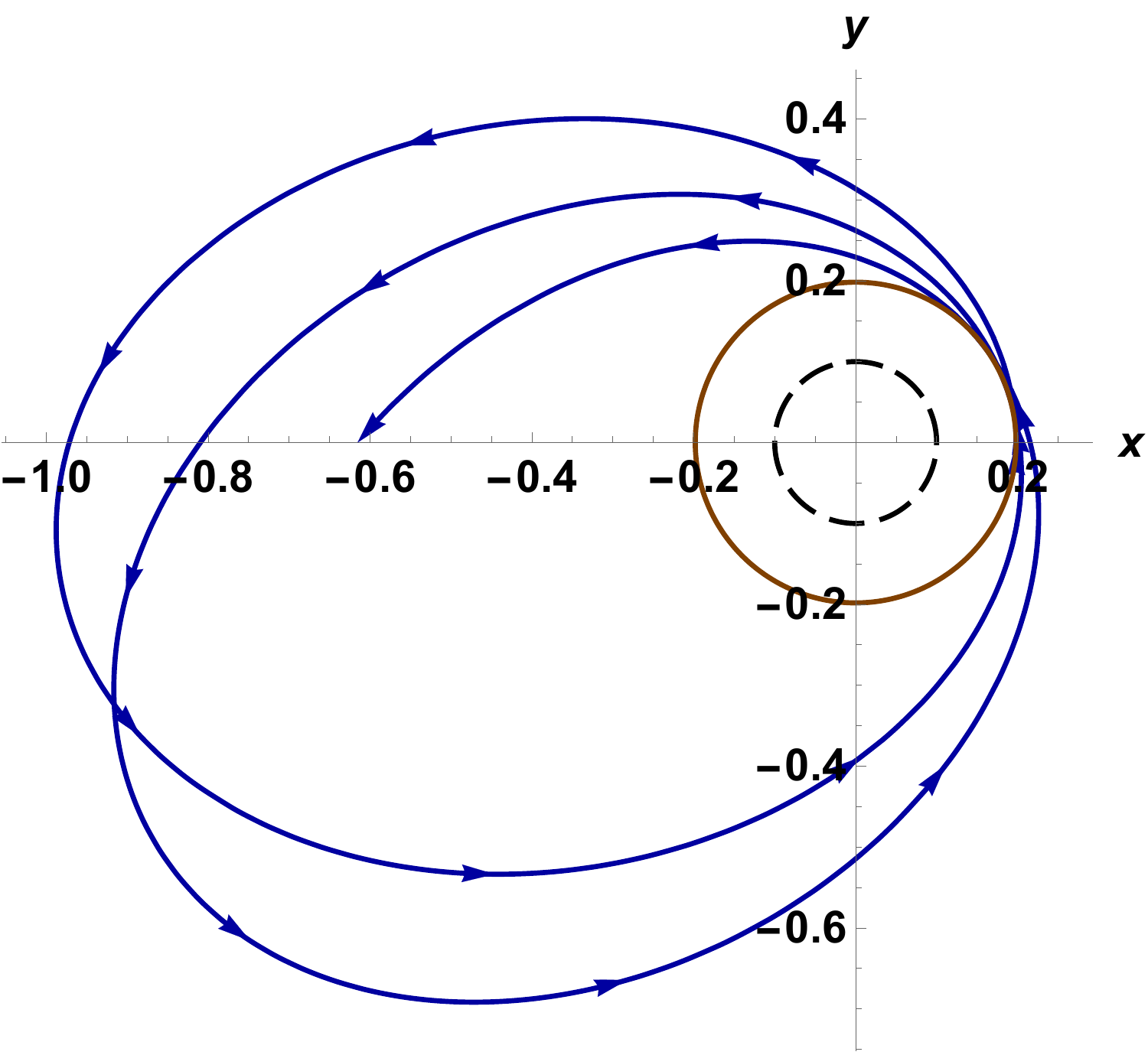}\label{orbitBSTM01}}
 \caption{In these figures, effective potential and particle orbits in Schwarzschild and BST spacetimes are shown. Here, we use a red line for Schwarzschild, while blue for BST. From the (\ref{orbitschM05},\ref{orbitBSTM05},\ref{orbitschM01},\ref{orbitBSTM01}), we can see that to reach one perihelion point to the other perihelion point, the angular distance travelled by a particle is greater than $2\pi$. The black dark spots in the figure (\ref{orbitschM05},\ref{orbitschM01}), show the black hole regions. The brown circles show the minimum approach of the particle near the center.}
 \label{precgen}
\end{figure*}

\begin{figure*}
\subfigure[$M_0 =0.1$, $h=0.2$, $E=-0.022$, $R_b=1$]
{\includegraphics[width=85mm]{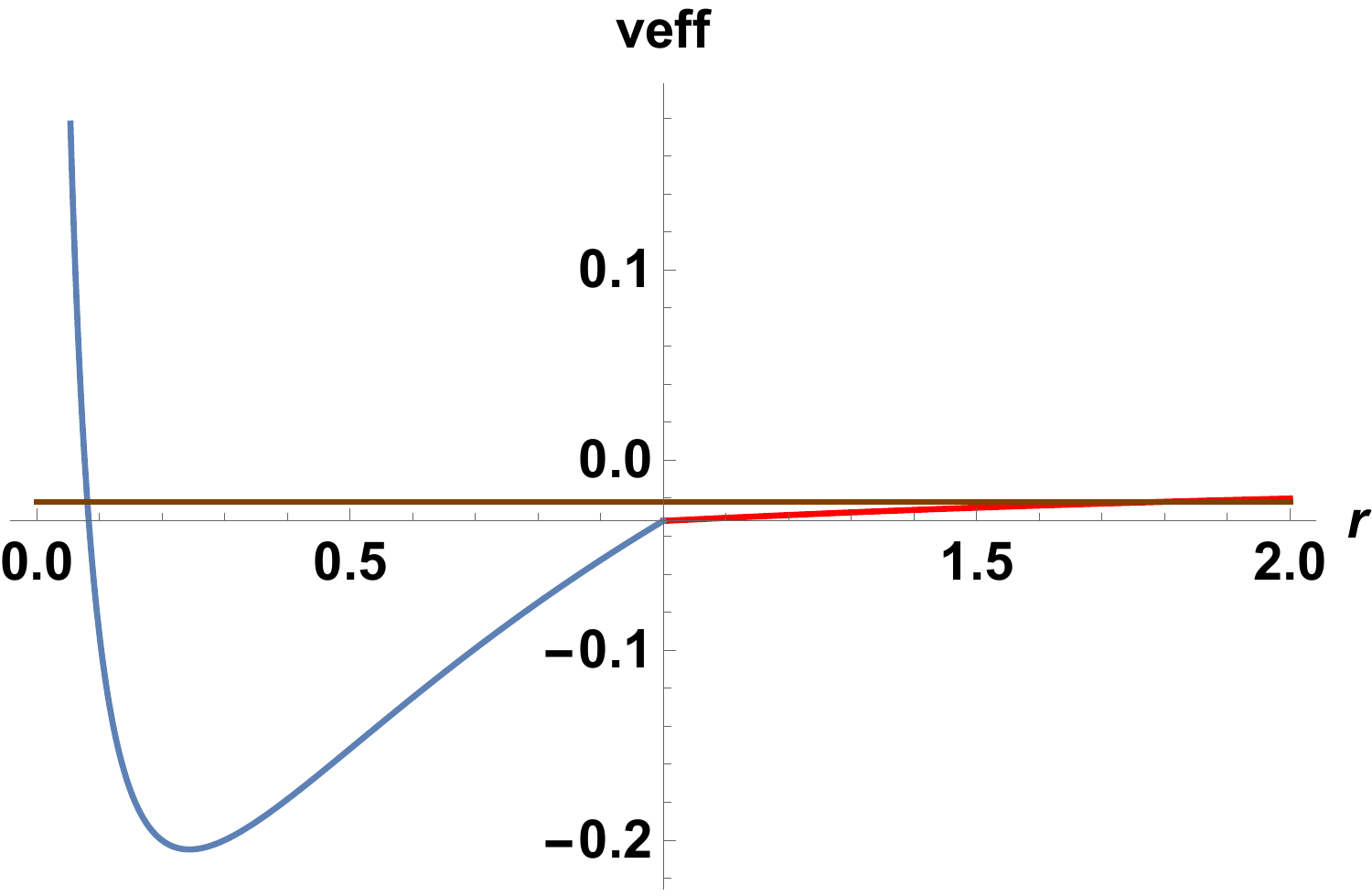}\label{G7}}
\subfigure[orbit in matching of BST with Schwarzschild spacetimes]
{\includegraphics[width=85mm]{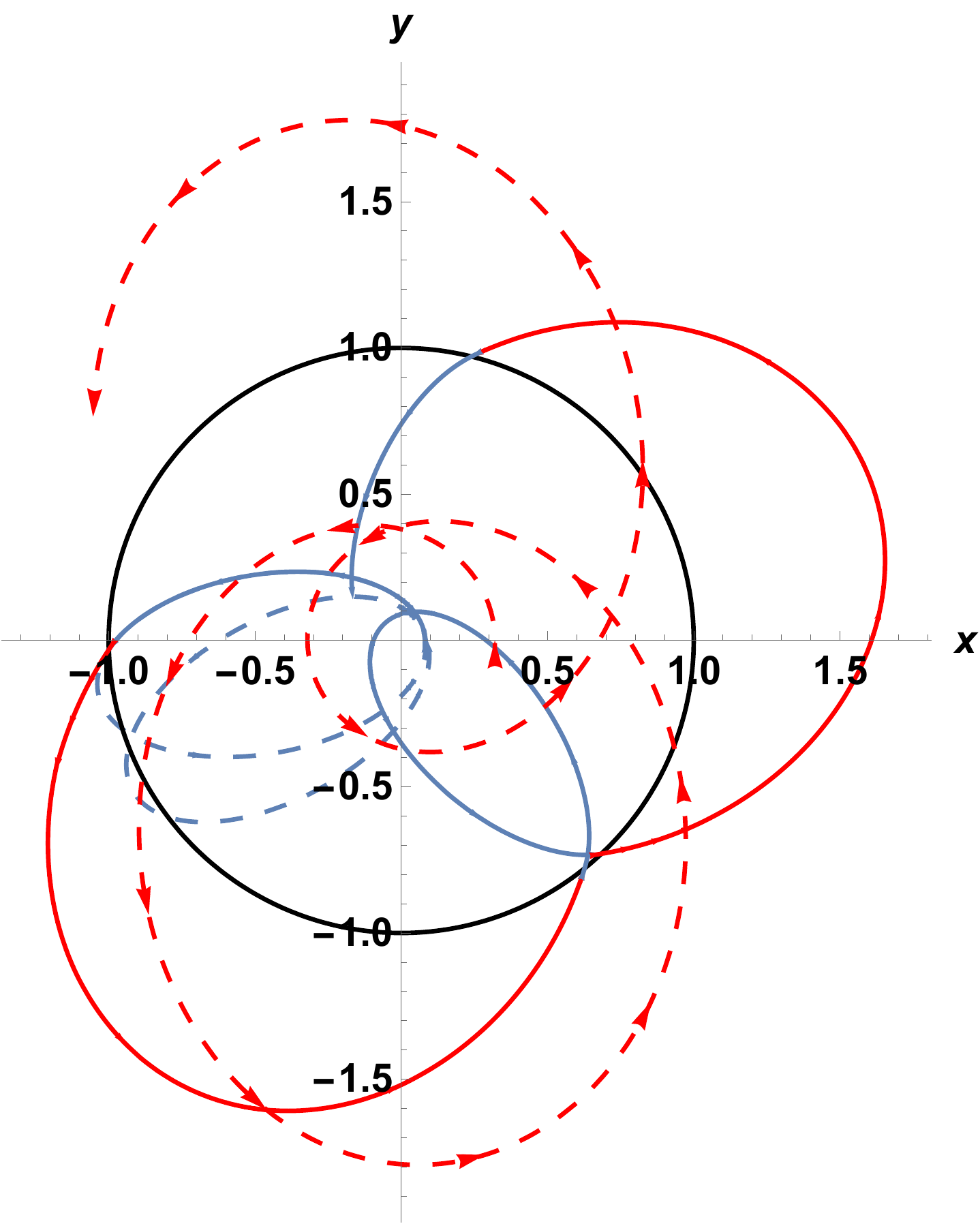}\label{H8}}
\subfigure[$M_0 =0.333$, $h=0.4$, $E=-0.044$, $R_b=1$]
{\includegraphics[width=85mm]{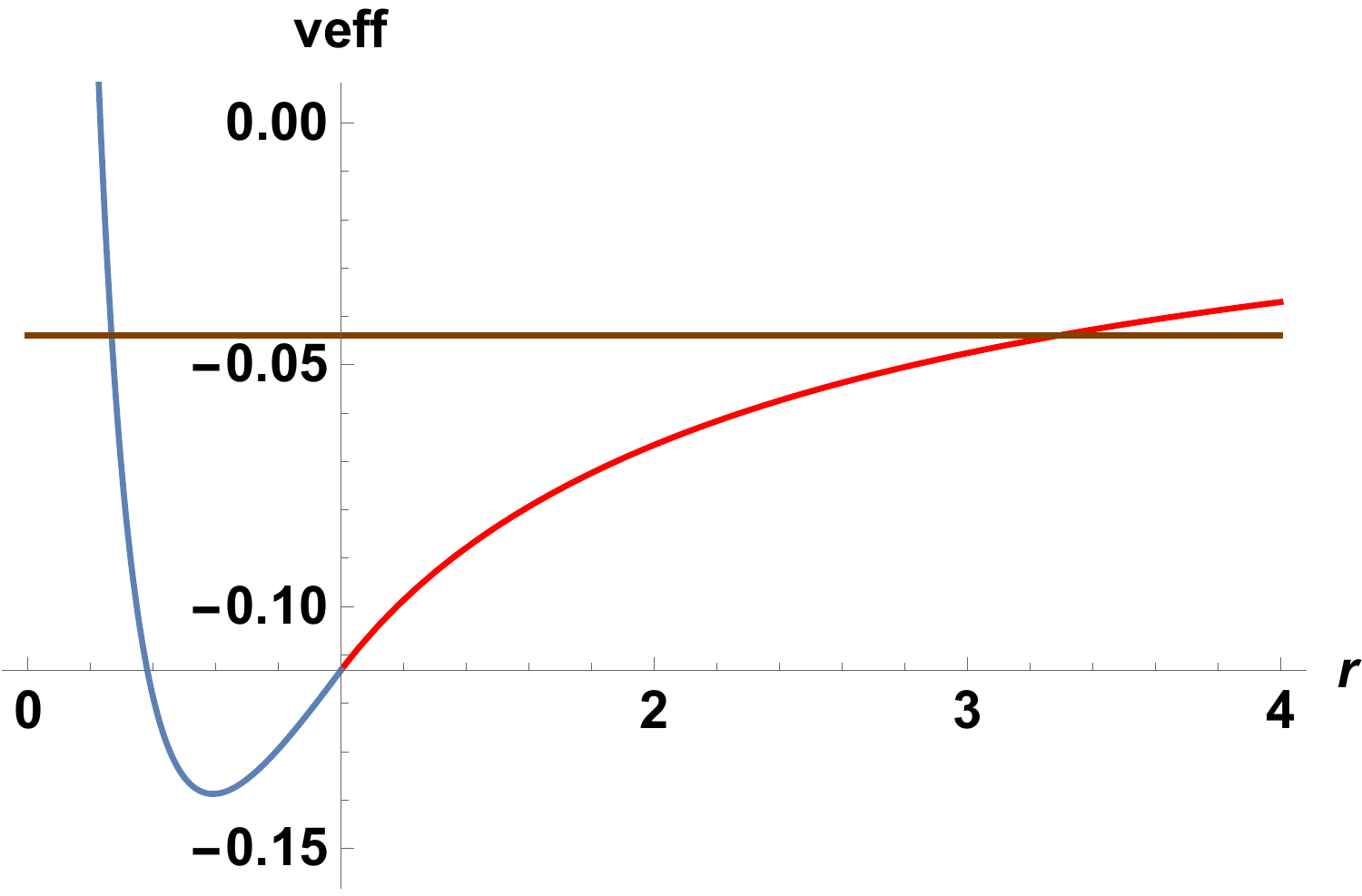}\label{G77}}
\subfigure[orbit in matching of BST with Schwarzschild spacetimes]
{\includegraphics[width=92mm]{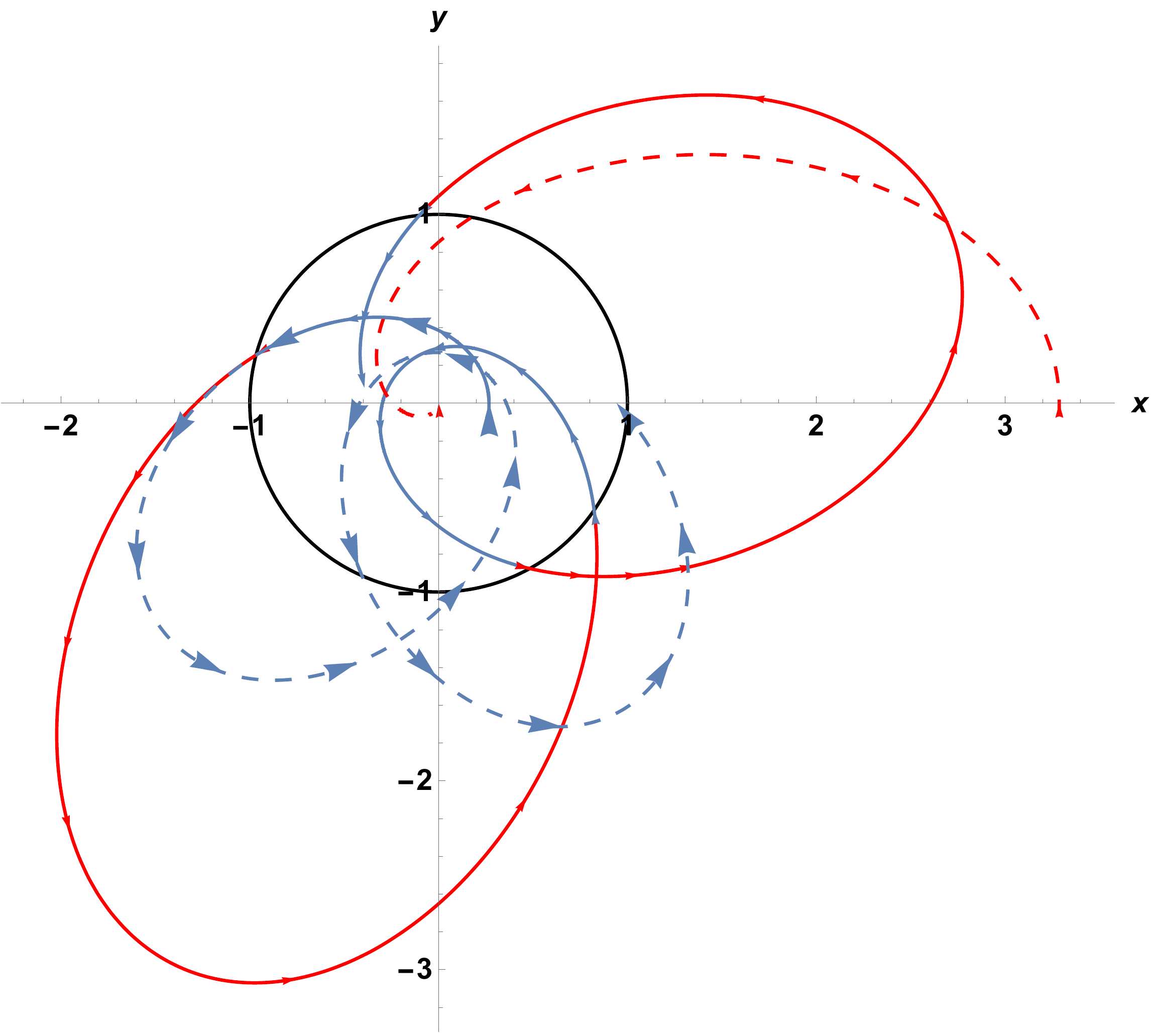}\label{H88}}
 \caption{In this figure, we show the nature of the orbit of a particle (fig.~(\ref{H8},\ref{H88})) and corresponding effective potential (fig.~(\ref{G7},\ref{G77})) when it crosses the matching radius $R_b$.    In the fig.~(\ref{H8},\ref{H88}) the dotted blue line shows what would be the particle trajectory if there is no Schwarzschild spacetime outside and dotted red line shows what would be the particle orbit when there is no BST spacetime inside. On the other
hand, the solid blue line shows particle’s actual path in  BST spacetime and the solid red line shows particle’s orbit in Schwarzschild spacetime. The solid red and blue part of effective potential shows the contribution of exterior Schwarzschild and interior BST spacetimes respectively. }
\end{figure*}

\section{The Orbit equation in Janis-Newman-Winicour (JNW) spacetime}
\label{JNWorbit}
The JNW spacetime is static, spherically symmetric and massless scalar field solution of Einstein equations \cite{JNW}. The Lagrangian density of minimally coupled scalar field can be written as,
\begin{equation}
\mathcal{L}=\sqrt{-g}\left(\frac12\partial^{\mu}\Phi\partial_{\mu}\Phi-V(\Phi)\right)\,\, ,
\end{equation}
with the minimal conditions: $R_{\mu\nu}-\frac12 R g_{\mu\nu}=\kappa T_{\mu\nu}$, $\Box\Phi(r)=V^{\prime}(\Phi(r))$, where the $\Phi$ is the scalar field, $R, R_{\mu\nu}, T_{\mu\nu}$ are the Ricci scalar, Ricci tensor, and energy-momentum tensor respectively, and $V(\Phi)$ is the scalar field potential and $\kappa$ is a constant whose value can be considered as 1 for $G=c=1$ unit system. The Energy-momentum tensor ($T_{\mu\nu}$) can be written as, $T_{\mu\nu}=\partial_{\mu}\Phi\partial_{\nu}\Phi-g_{\mu\nu}\mathcal{L}$. JNW spacetime is an example of this minimally coupled scalar field which has zero mass. This spacetime is asymptotically flat, therefore, we need not match theJNW spacetime with Schwarzschild spacetime. The line element  can be written as,
\begin{figure*}
\centering
\subfigure[$M =0.025$,$q=0.06$,$h=0.1$,$E=-0.02$]
{\includegraphics[width=65mm]{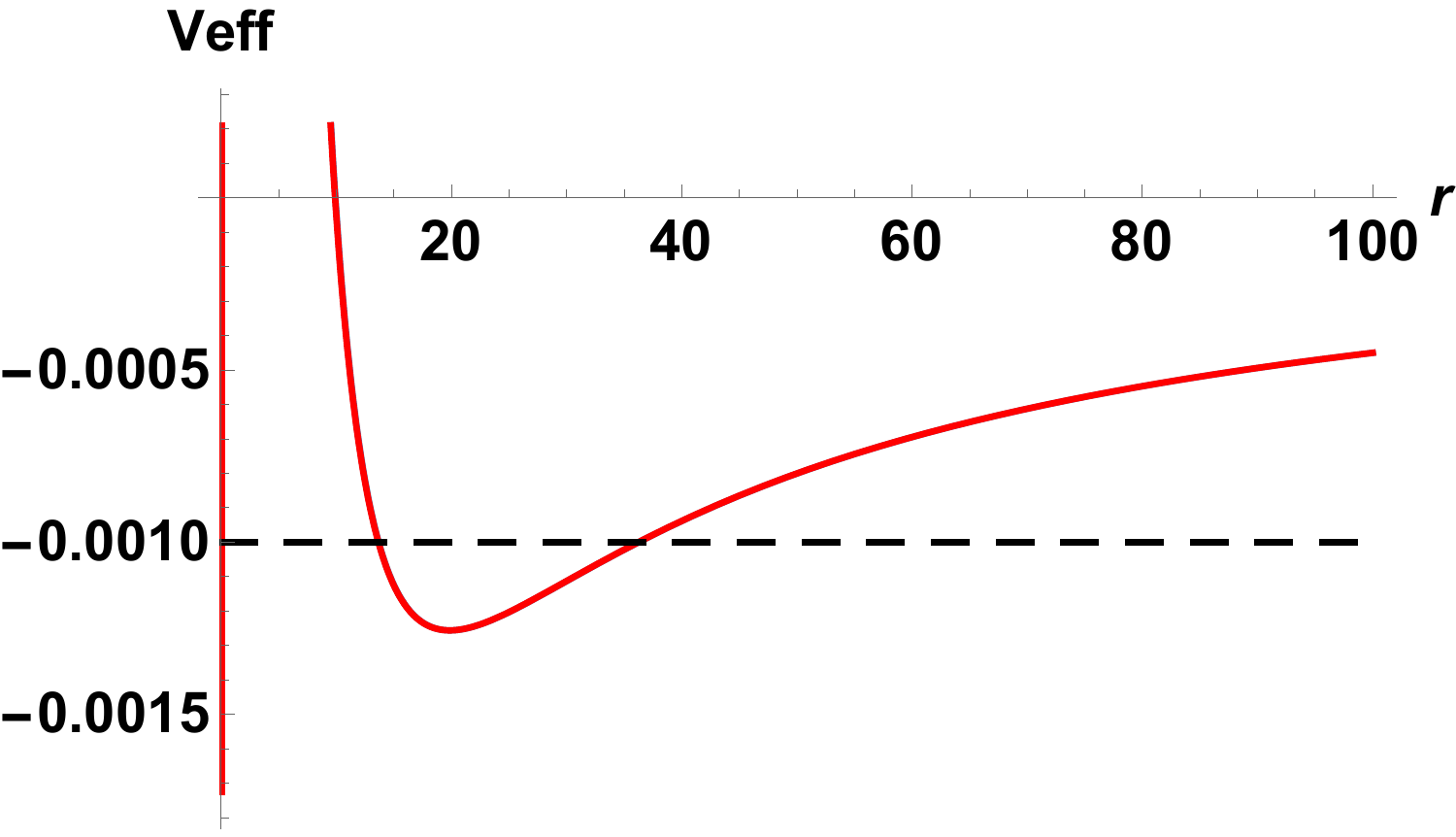}\label{VeffJNWq01}}
\hspace{0.2cm}
\subfigure[Particle orbit in Schwarzschild spacetime]
{\includegraphics[width=65mm]{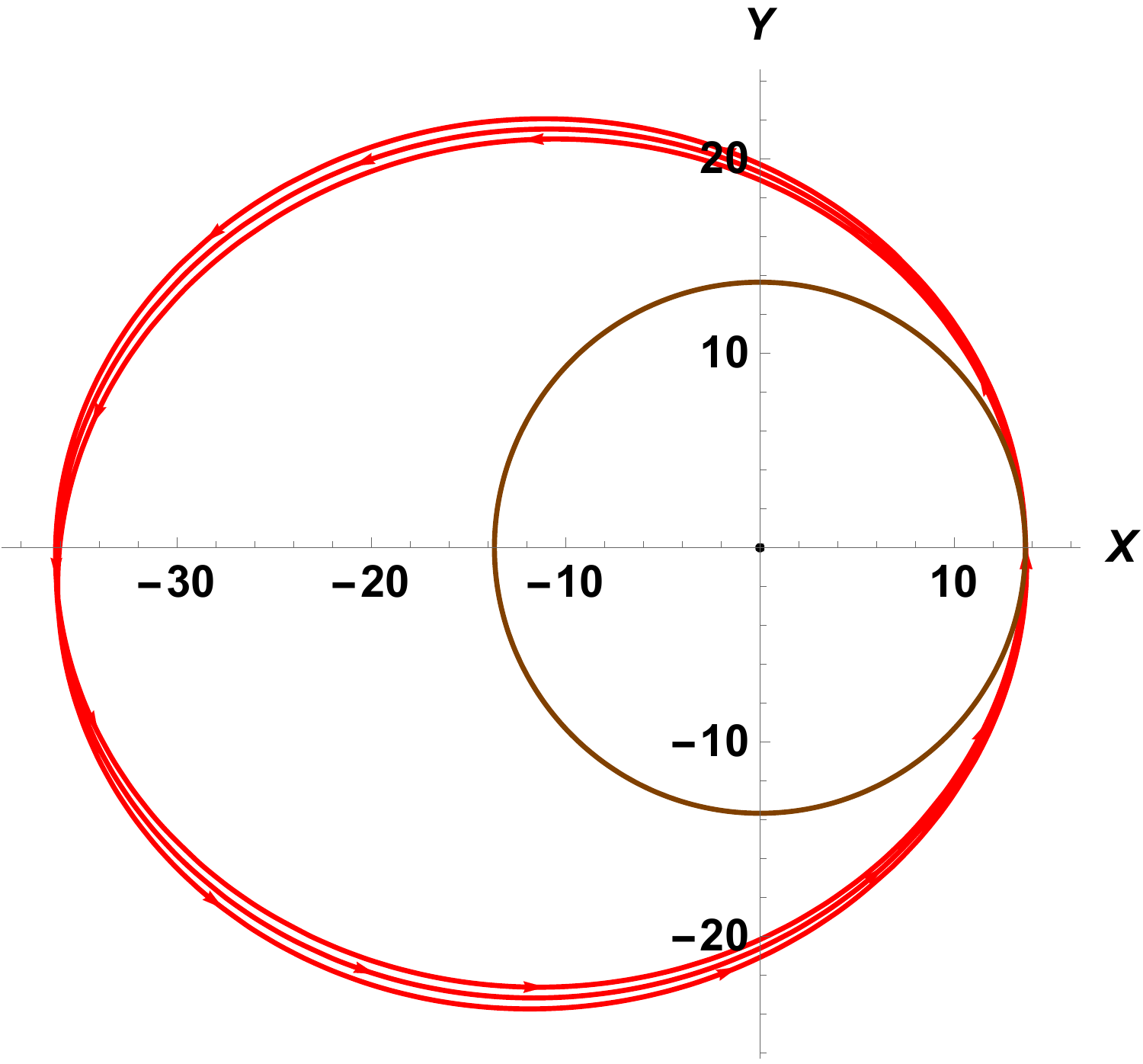}\label{orbitschq01}}
\hspace{0.2cm}
\subfigure[Particle orbit in JNW spacetime]
{\includegraphics[width=65mm]{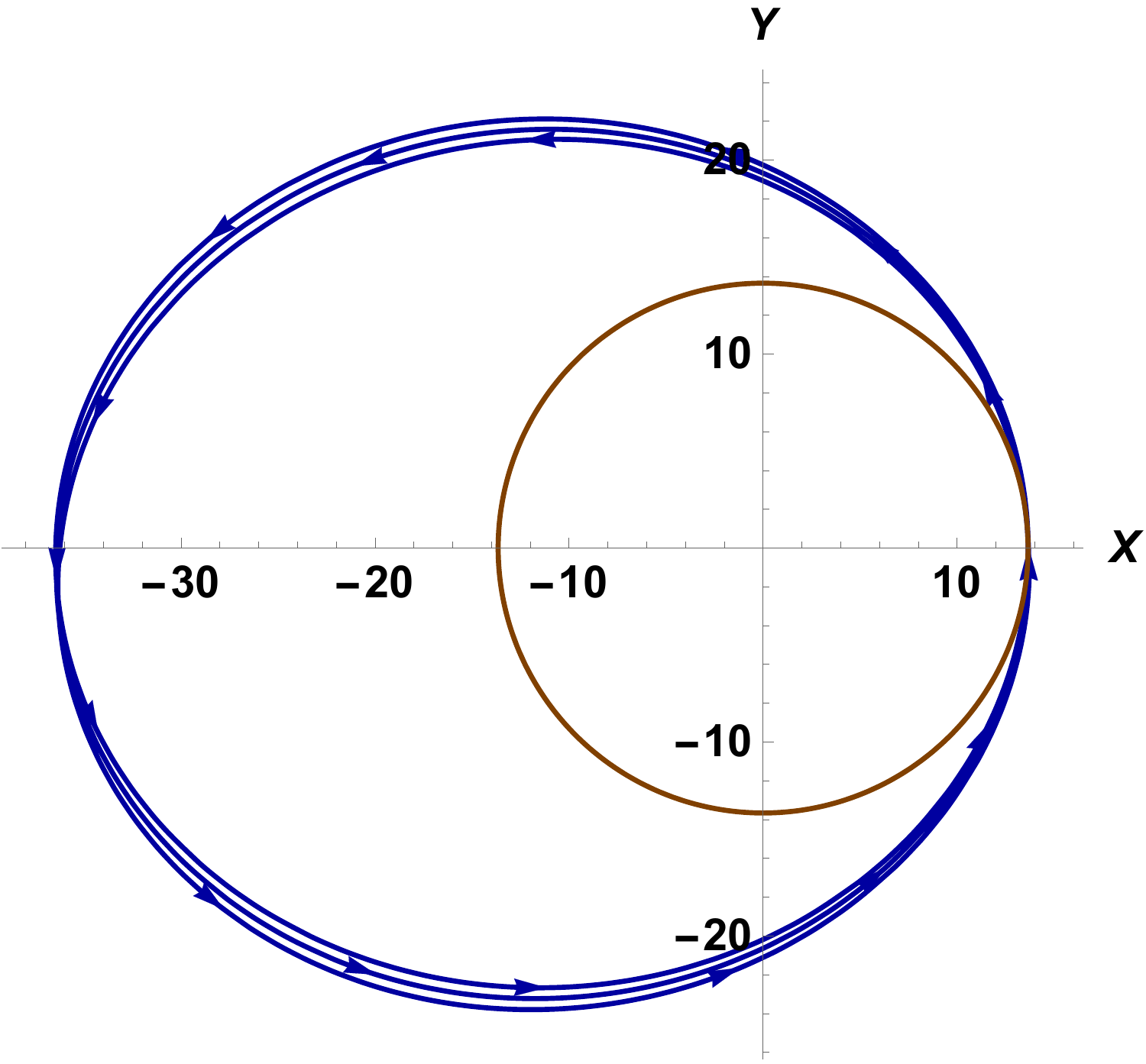}\label{orbitJNWq01}}
\subfigure[$M =0.025$,$q=0.5$,$h=0.1$, $E=-0.02$]
{\includegraphics[width=80mm]{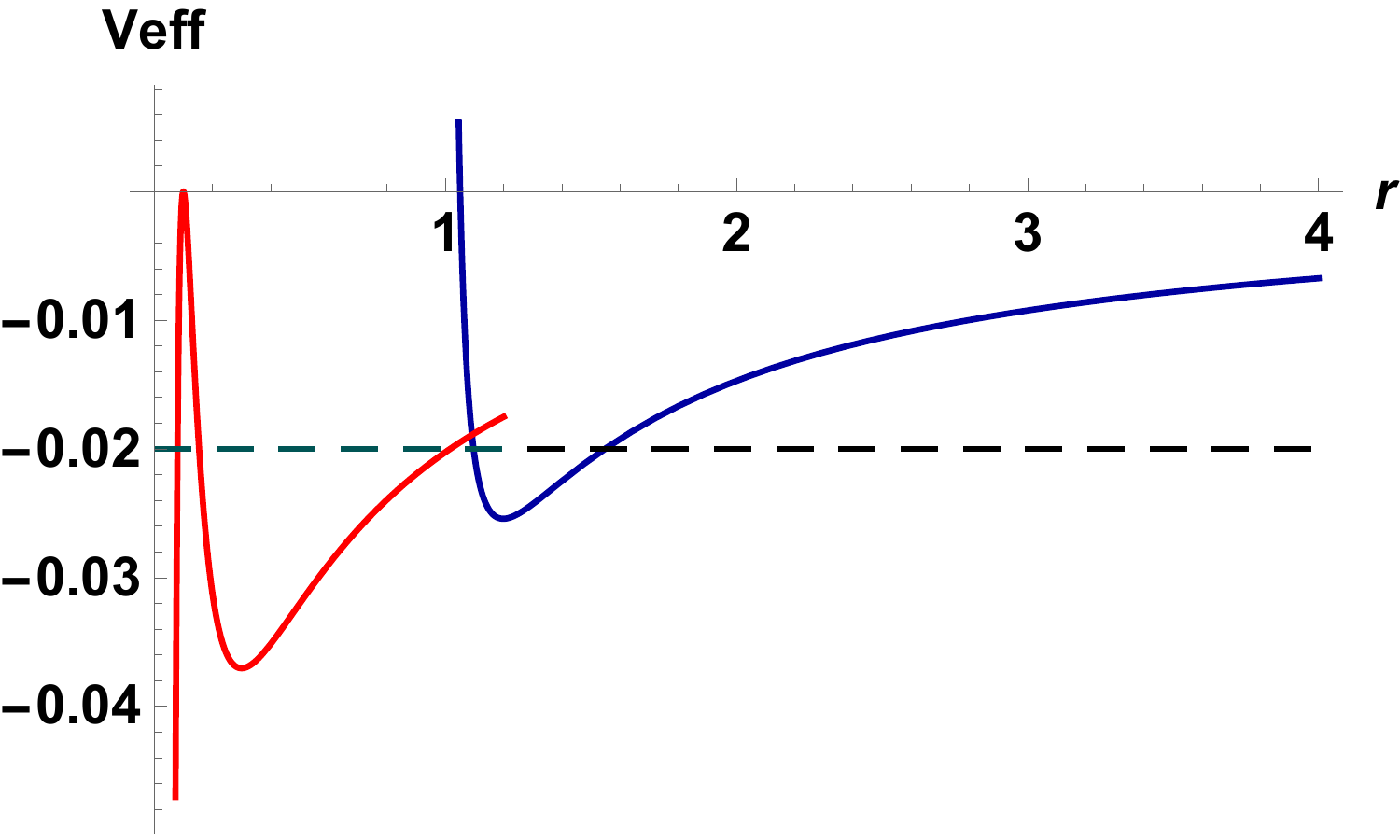}\label{VeffJNWq05}}
\subfigure[Particle orbit in Schwarzschild spacetime]
{\includegraphics[width=65mm]{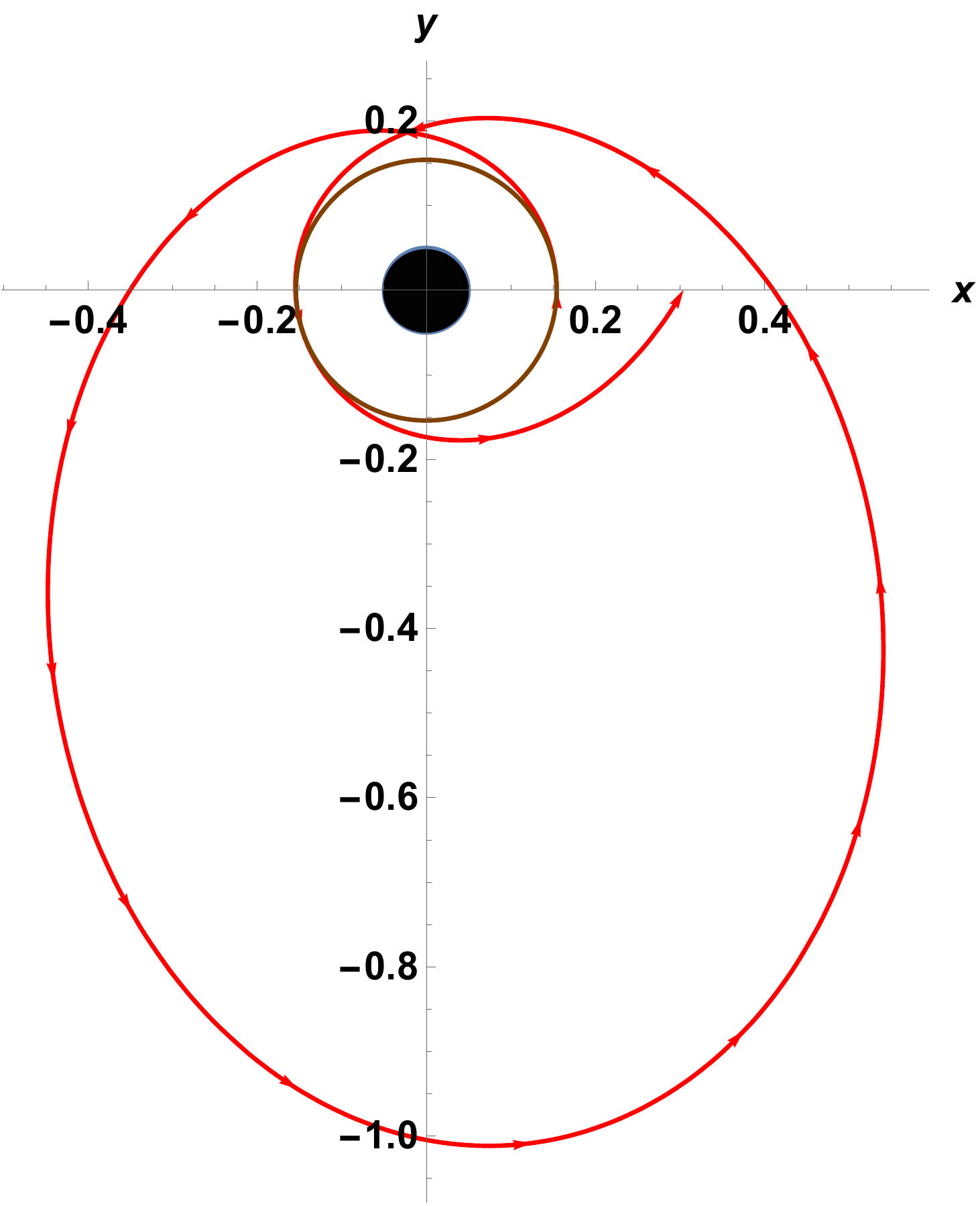}\label{orbitschq05}}
\hspace{0.2cm}
\subfigure[Particle orbit in JNW spacetime]
{\includegraphics[width=65mm]{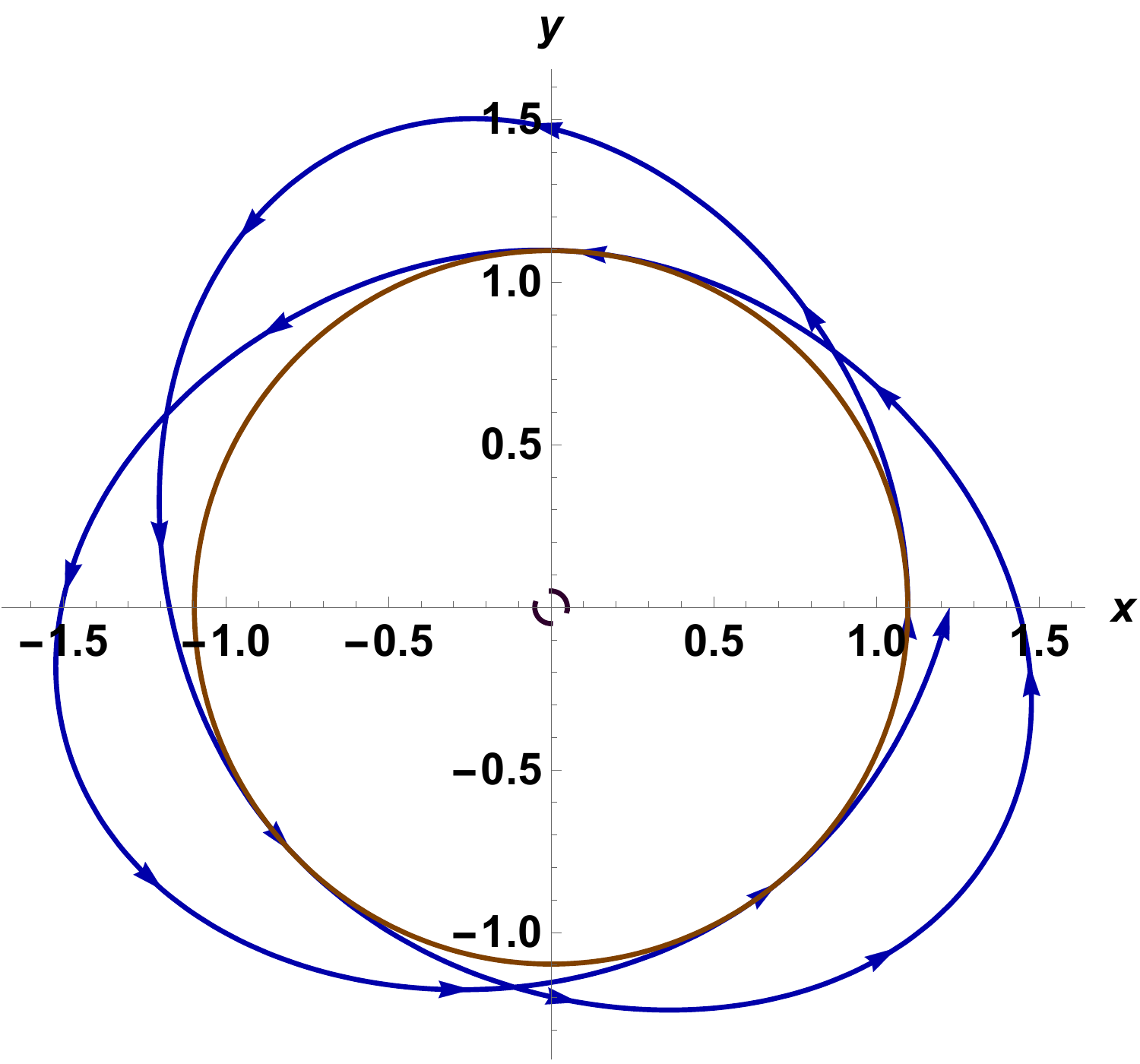}\label{orbitJNWq05}}
 \caption{In this figure, we show the effective potentials (fig.~(\ref{VeffJNWq01},\ref{VeffJNWq05})) and the orbits of particles in Schwarzschild and JNW spacetimes (fig.~(\ref{orbitJNWq01},\ref{orbitJNWq05}, \ref{orbitschq01},\ref{orbitschq05})) for $M=0.025, q=0.1$ and $q=0.5$. In this diagram, the blue line and red line correspond to JNW and Schwarzschild spacetimes respectively and the dotted horizontal black line is indicating the total energy of the freely falling particle. The brown circle shows the perihelion positions of a particle near the center and dark black region at the center in diagrams (fig.~(\ref{orbitschq01},\ref{orbitschq05})), shows the positions of the black holes. Here we consider $R_b=1$.}
\label{M0025} 
\end{figure*}
\begin{widetext}
\begin{eqnarray}
 ds^2_{JNW} &=& -\left(1-\frac{b}{r}\right)^n dt^2 + \left(1-\frac{b}{r}\right)^{-n}dr^2 + r^2\left(1-\frac{b}{r}\right)^{1-n}d\Omega^2\,\, ,
 \label{JNWmetric}
\end{eqnarray}
\end{widetext}
 \begin{figure*}
 \subfigure[$M =0.25$,$q=3$,$h=3$, $E=-0.002$]
 {\includegraphics[width=65mm]{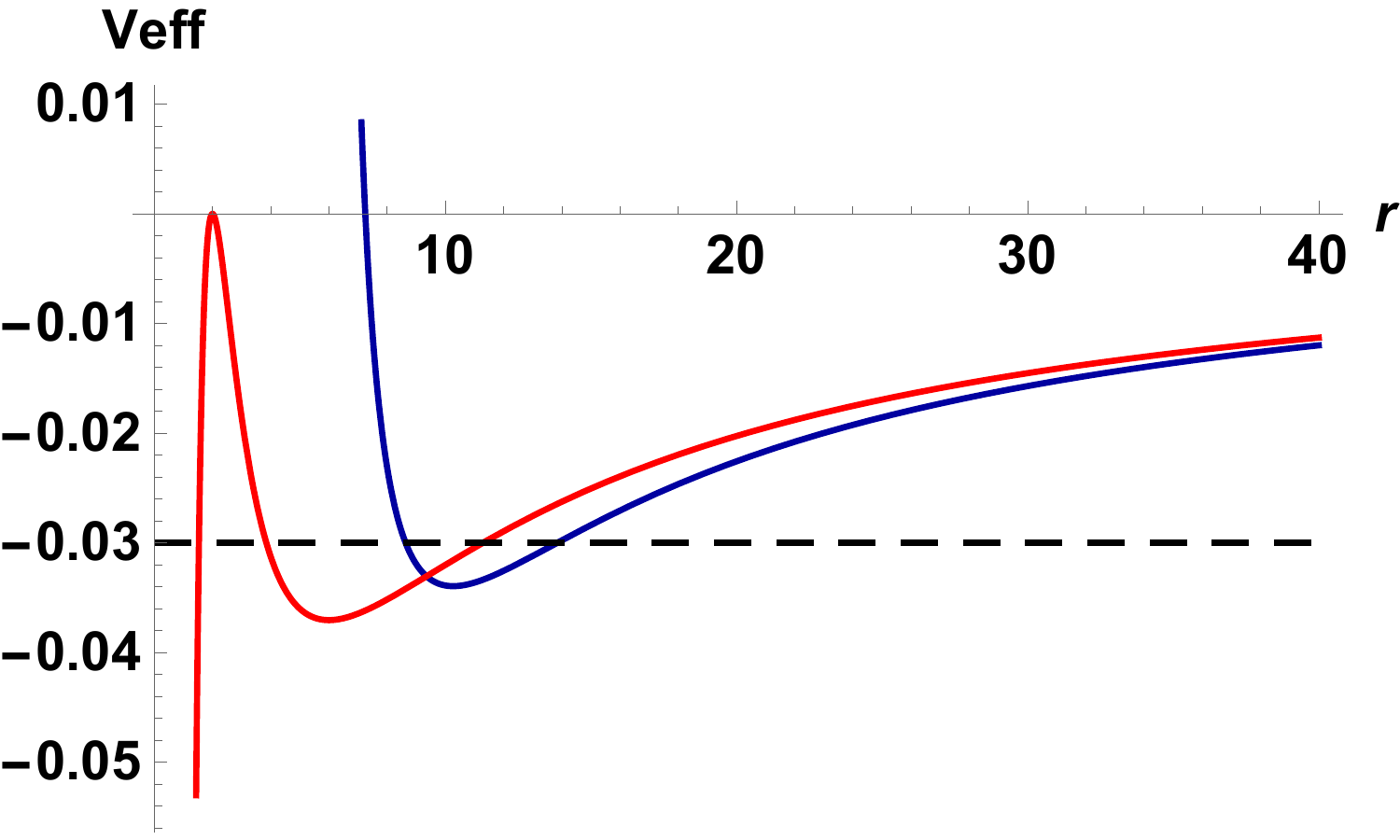}\label{VeffJNWq4}}
 \hspace{0.4cm}
 \subfigure[Particle orbit in Schwarzschild spacetime]
 {\includegraphics[width=65mm]{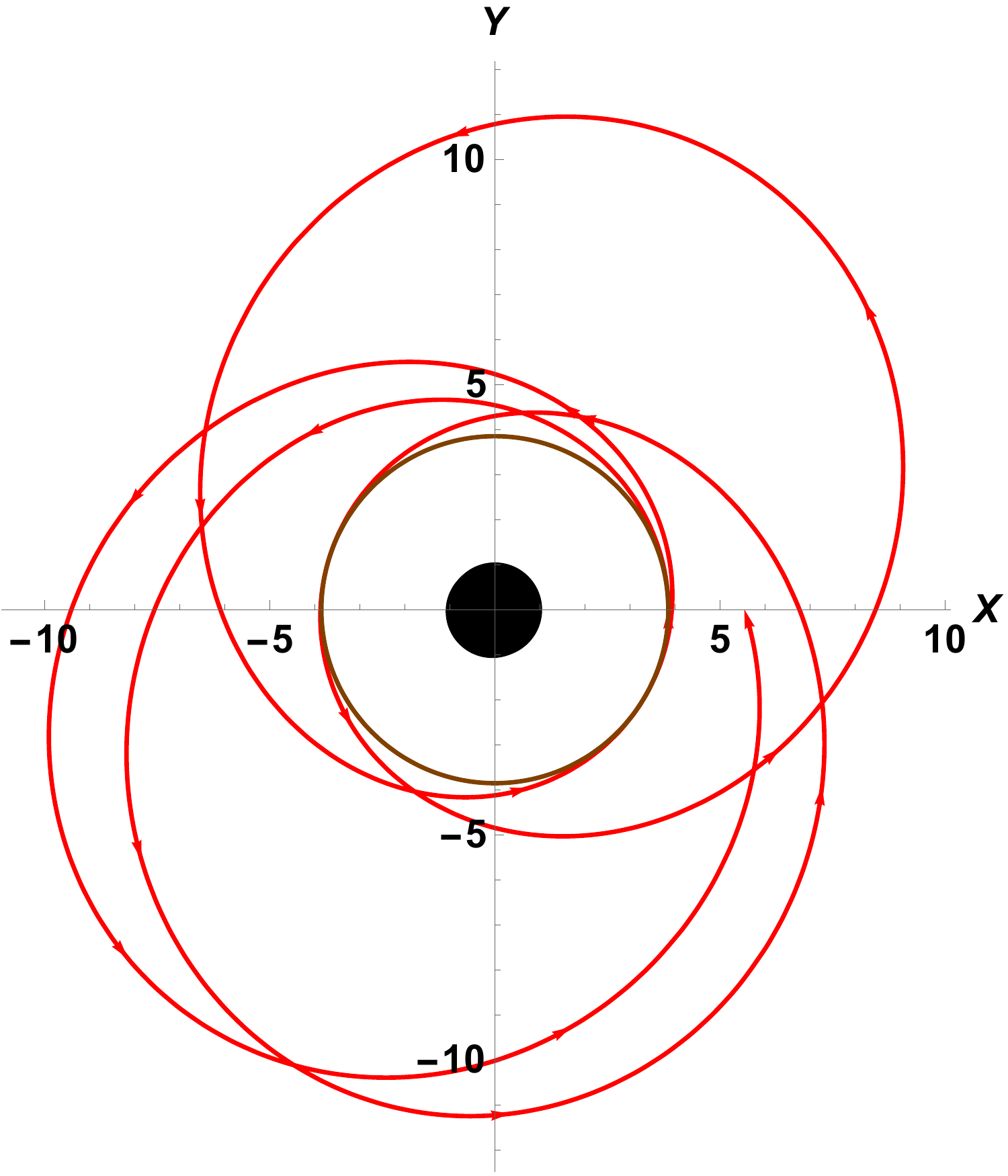}\label{orbitschq4}}
 \hspace{0.2cm}
 \subfigure[Particle orbit in JNW spacetime]
 {\includegraphics[width=65mm]{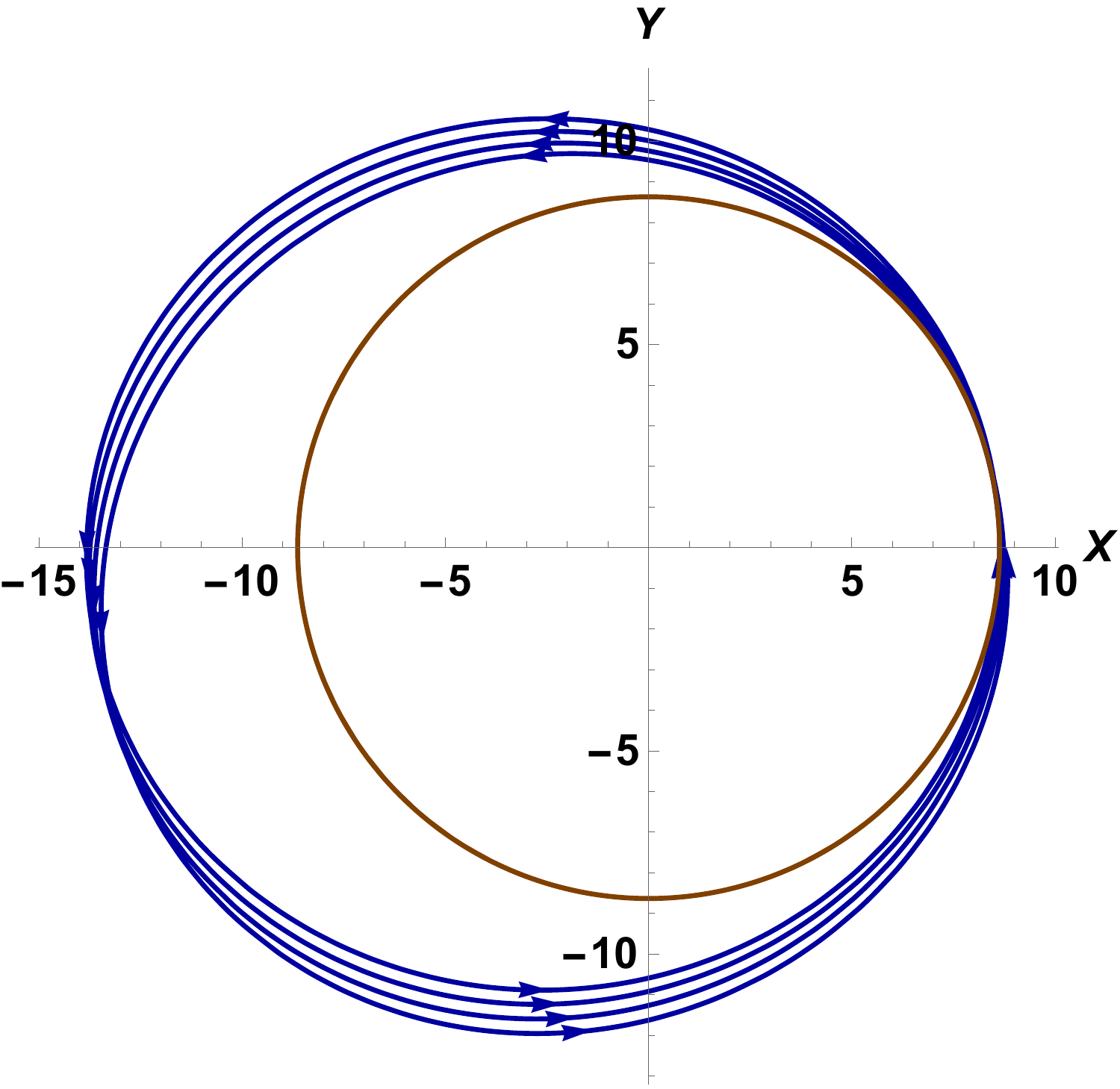}\label{orbitJNWq4}}
 \hspace{0.4cm}
 \subfigure[$M =0.25$,$q=10$,$h=3$, $E=-0.002$]
 {\includegraphics[width=65mm]{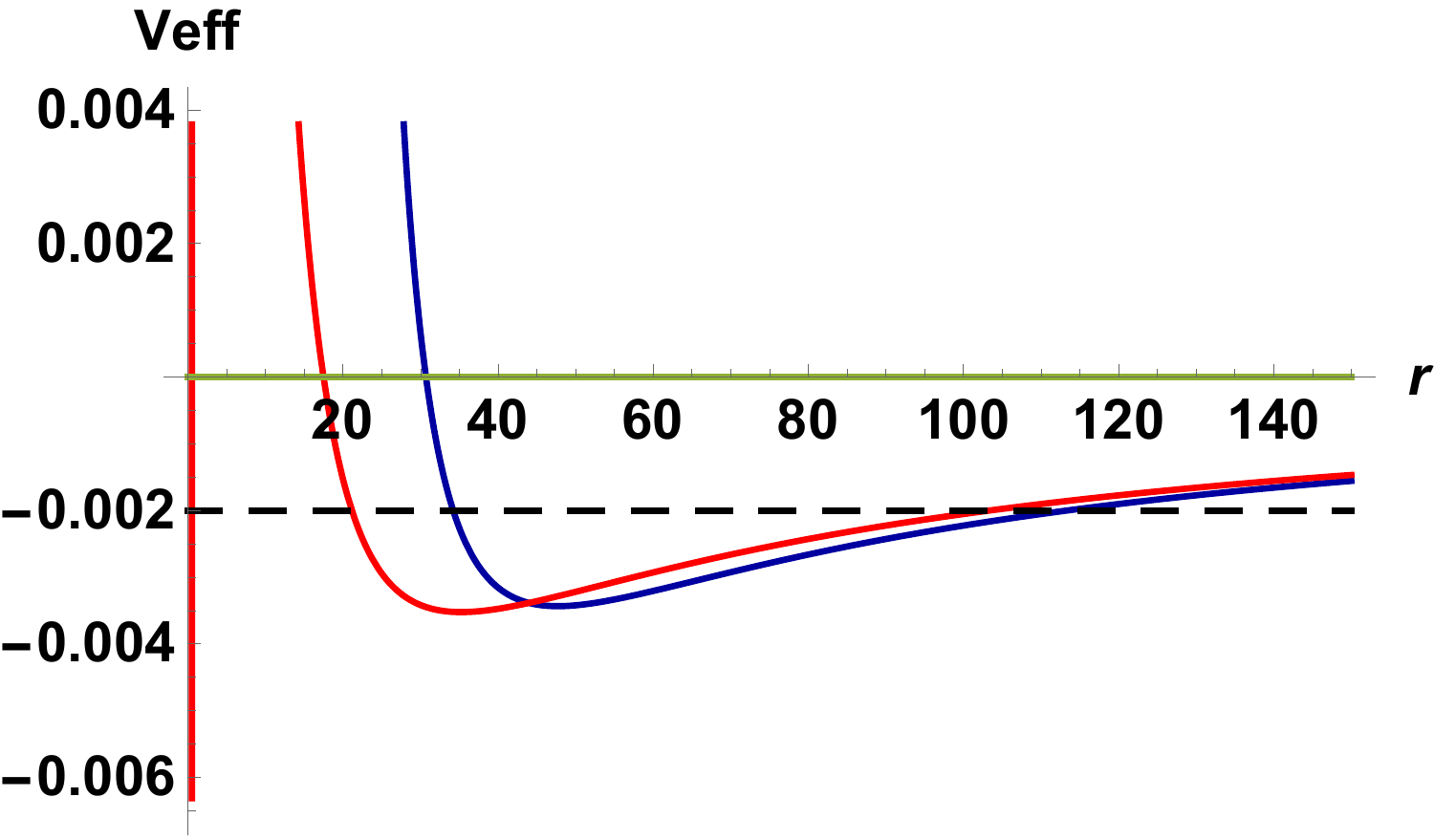}\label{VeffJNWq10}}
\hspace{0.2cm}
\subfigure[Particle orbit in Schwarzschild spacetime]
{\includegraphics[width=65mm]{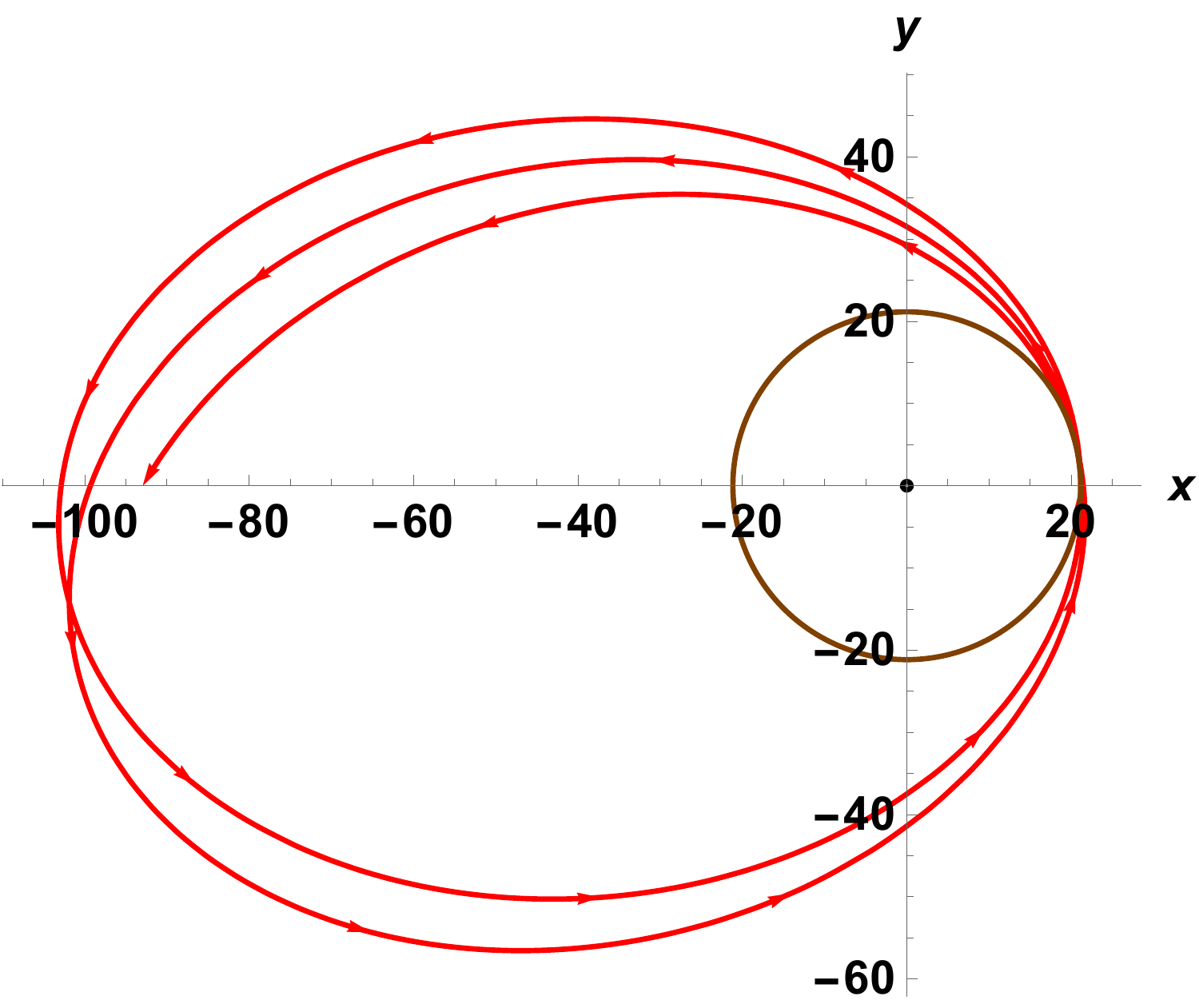}\label{orbitschq10}}
 \subfigure[Particle orbit in JNW spacetime]
 {\includegraphics[width=65mm]{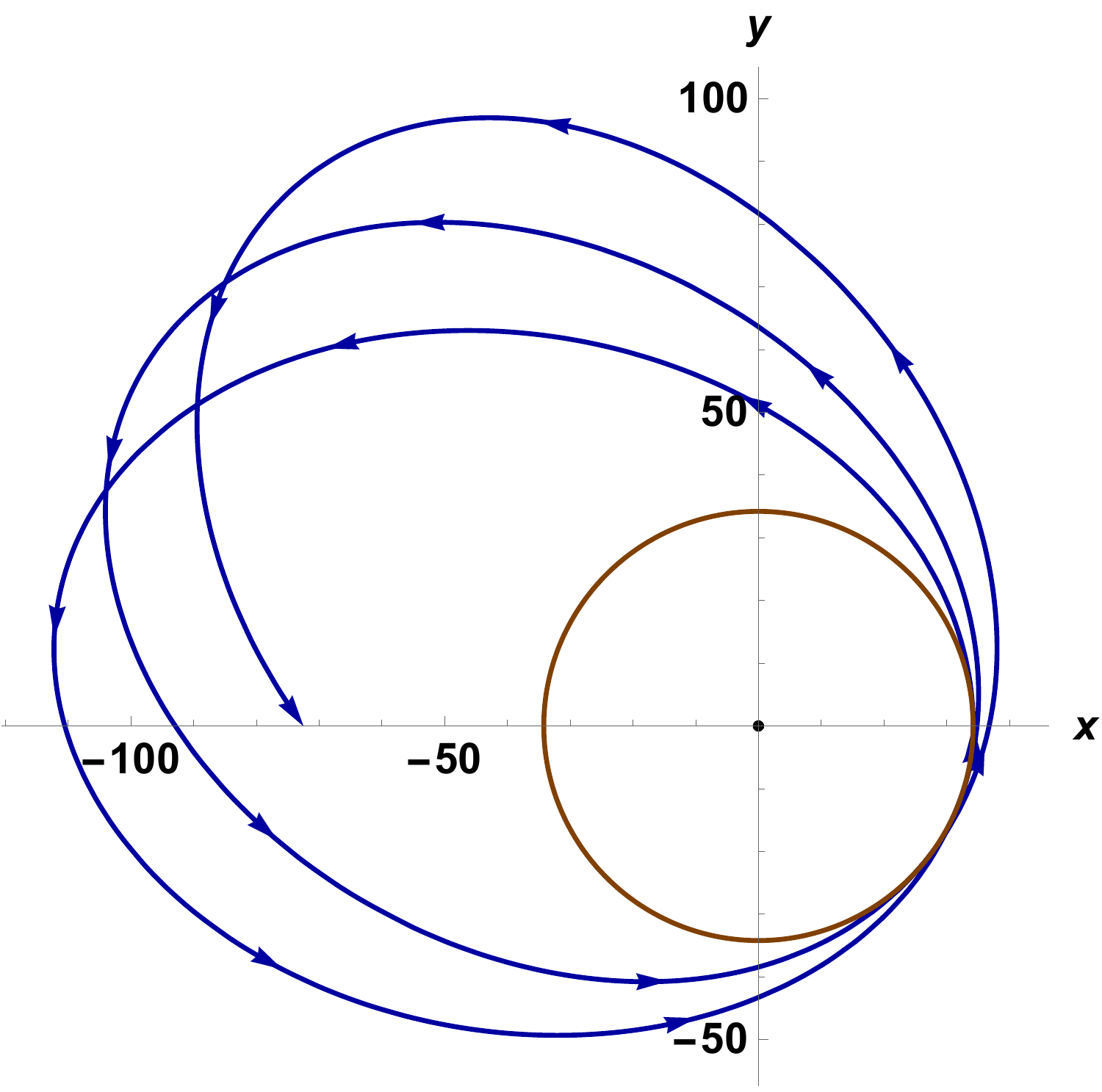}\label{orbitJNWq10}
}
\caption{
Here, the effective potential (fig.~(\ref{VeffJNWq4},\ref{VeffJNWq10})) and particle  orbits in Schwarzschild and JNW spacetimes (fig.~(\ref{orbitschq4},\ref{orbitschq10},\ref{orbitJNWq4},\ref{orbitJNWq10})) are shown for $M=0.25, q=3$ and $q=10$. It can be seen that for scalar field charge $q=3$, perihelion precession direction in JNW spacetime is Schwarzschild like, while, for scalar field charge $q=10$, perihelion precession direction in JNW spacetime is opposite to the direction of precession in the Schwarzschild spacetime. Here, again we consider $R_b=1$. }
\label{M025}
\end{figure*}
where, $b=2\sqrt{M^2+q^2}$ and $ n=\frac{2M}{b}$. $M$ and $q$ are the two constant parameters of JNW spacetime, which represent total mass and scalar field charge respectively. From the expressions of $b$ and $n$, we can write $0<n<1$. The massless scalar can be written as,
\begin{equation}
    \Phi=\frac{q}{b\sqrt{4\pi}}ln\left(1-\frac{b}{r}\right)\,\, .
\end{equation}
If we consider $n=1$ or $q=0$, there will not be any scalar field ($\Phi$) effects and JNW spacetime becomes Schwarzschild spacetime. As we know JNW spacetime is static and spherically symmetric, so that conserved angular momentum ($h_{JNW}$) and energy ($\gamma_{JNW}$) per unit rest mass of the freely falling particle in JNW spacetime are given by,
\begin{eqnarray}
h_{JNW} = r^2\left(1-\frac{b}{r}\right)^{1-n}\frac{d\phi}{d\tau}\, ,\,\,\nonumber\\
\gamma_{JNW} =  \left(1-\frac{b}{r}\right)^{n}\frac{dt}{d\tau}\,\,,
\label{conservedJNW}
\end{eqnarray}

where $\tau$ is the proper time of the particle and we consider $\theta=\frac{\pi}{2}$. From normalization of four velocity of a freely falling massive particle, we can write the following effective potential for JNW spacetime, 
\begin{widetext}
\begin{eqnarray}
(V_{eff})_{JNW} &=& \frac{1}{2}\left[\left(1-\frac{b}{r}\right)^n\left(1+\frac{h_{JNW}^2}{r^2}\left(1-\frac{b}{r}\right)^{n-1}\right)-1\right]\,\, .
\end{eqnarray}
\end{widetext}
Using the condition for stable circular orbits, we can write down the expressions of $h$ and $\gamma$ for JNW spacetime as,
\begin{eqnarray}
\gamma_{JNW}^2 &=& \left(1-\frac{b}{r}\right)^n\left[\frac{2r-b(n+1)}{2r-b(2n+1)}\right]\,\, ,\\
h_{JNW}^2 &=& r^2\left[\frac{bn\left(1-\frac{b}{r}\right)^{1-n}}{2r-b(2n+1)}\right]\,\, .
\end{eqnarray}

For stability of circular orbits, as we know, we need $(V^{\prime\prime}_{eff})>0$. However, for JNW spacetime, $(V^{\prime\prime}_{eff})_{JNW}$ is not always positive \cite{Gyulchev:2019tvk},\cite{Zhou:2014jja}. For $n<0.447$, at any point in JNW spacetime, we have $(V^{\prime\prime}_{eff})_{JNW}>0$ and, therefore a freely falling particle can have stable, circular orbit at any point in JNW spacetime. However, for $0.447<n<0.5$, there exists one certain radial interval ($r_{-},r_{+}$) inside which no stable circular orbits are possible, where $r_{-}$ and $r_{+}$ can be written as,
\begin{eqnarray}
r_{-}&=&\frac14\left(b~(2+6n)-4.472~b\sqrt{n^2-0.2}\right)\,\, ,\label{osco}\\
r_{+}&=&\frac14\left(b~(2+6n)+4.472~b\sqrt{n^2-0.2}\right)\,\, .
\label{isco1}
\end{eqnarray}
In $r_{-}<r<r_{+}$, we always have $(V^{\prime\prime}_{eff})_{JNW}<0$ and, therefore, stable circular orbits are not possible in that interval of radial distance. However, outside this radial interval stable circular orbits are possible, as there we have $(V^{\prime\prime}_{eff})_{JNW}>0$. Now, for $0.5<n<1$, there exists the inner most circular orbit at $r_{ISCO}$, below which no stable circular orbits are possible. Therefore, unlike BST spacetime, JNW spacetime can have ISCO for $0.5<n<1$.   
Using general method of finding orbit equation \cite{Parth}, one can write the orbit equation of a particle freely falling in JNW spacetime as,
\begin{widetext}
\begin{eqnarray}
   \frac{d^2u}{d\phi^2} + u + \frac{b\gamma^2}{2h^2}(2-2n)(1-bu)^{1-2n} - \frac{b(2-n)}{2h^2}(1-bu)^{1-n} - \frac{3bu^2}{2}&=&0\,\,
   \label{orbiteqJNW}
\end{eqnarray}
\end{widetext}
\begin{figure*}[t!]
\centering
\subfigure[$h=3$,$E=-0.002, R_b=1$]
{\includegraphics[width=75mm]{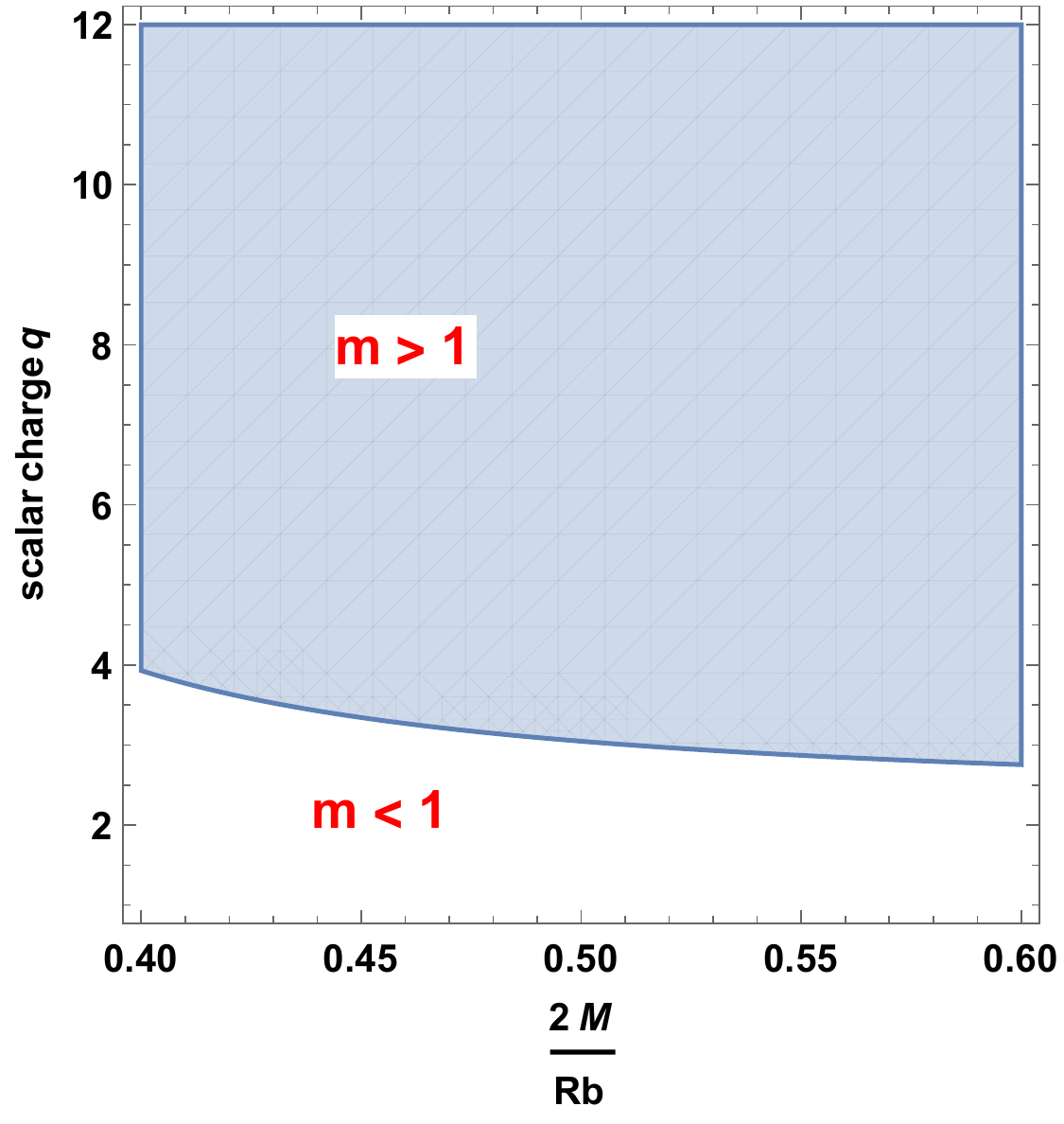}\label{re1}}
\hspace{0.2cm}
\subfigure[$h=0.1$,$E=-0.02, R_b=1$]
{\includegraphics[width=77mm]{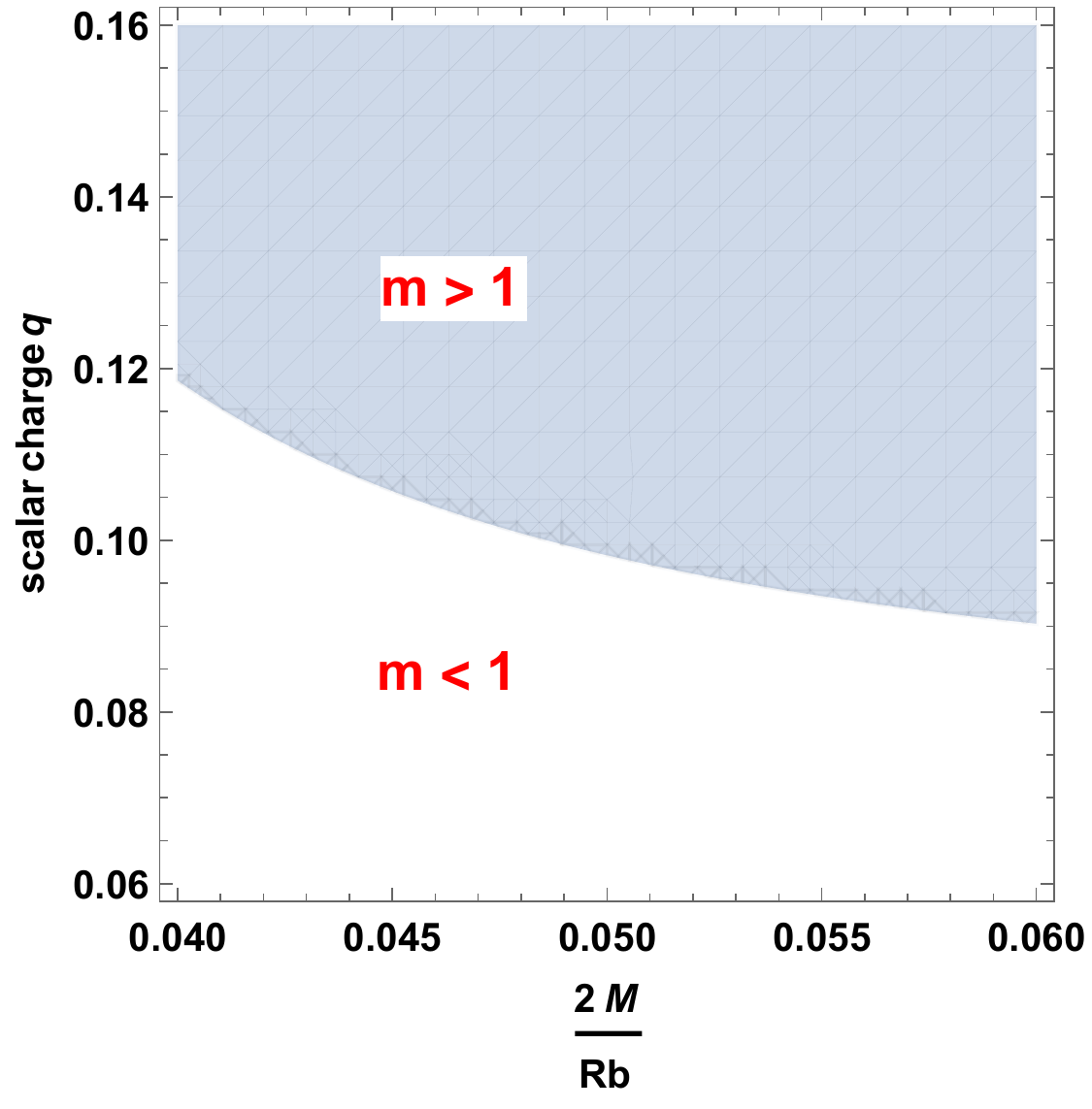}\label{re2}}
 \caption{Here, we show the regions of $m>1$ and $m<1$ in  the parameter space of the scalar charge $q$ and ADM mass $M$ of the JNW spacetime.}
\end{figure*}

\section{Approximate solution of orbit equations in Schwarzschild, BST and JNW spacetimes}\label{approximation}
The approximate solution of the orbit equations can be obtained by considering the small values of eccentricity ($e$). In \cite{Parth, Struck:2005hi, Struck:2005oi}, this method is extensively discussed. With this approximation, one can get important information about the nature and shape of bound orbits which are difficult to get from orbit equations. 
 

We can write the approximate solution for the orbit eq.~(\ref{orbiteqsch}) of Schwarzschild spacetime,
\begin{equation}
\tilde{u}=\frac{1}{p}\left[1+e\cos(m\phi)+O(e^2)\right]\,\, ,
\label{orbitsch1}
\end{equation}
where $m$ and $p$ are positive real numbers and $\tilde{u}=R_b u$. As we know, the solution of Newtonian orbit equation can be written as $\tilde{u}=\frac{1}{p}\left[1+e\cos(\phi)\right]$, therefore, with the above approximation we can get modified expression of $p$ and $m$. In Newtonian mechanics, $p=\frac{2h^2}{M_0R_b^2}$ and $m=1$. Therefore, perihelion precession cannot be possible in Newtonian mechanics. Using the approximate solution of orbit eq.~(\ref{orbitsch1}), we can get the following expression of $p$ and $m$ for Schwarzschild spacetime,
\begin{eqnarray}
   p=\frac{1+\sqrt{1-\frac{3M_0^2R_b^2}{h^2}}}{\frac{M_0R_b^2}{h^2}}\,\, ,\label{psch}\\
   m=\sqrt{1-\frac{3M_0}{p}}\label{msch}\,\, ,
\end{eqnarray}
where, $0<m<1$ and $m>1$ implies that, starting from a perihelion point, a particle reaches another perihelion point after travelling an angular distance of greater or less than $2\pi$ respectively. For Schwarzschild case, $m$ is always less than one \cite{Parth} and therefore, in Schwarzschild spacetime, a particle has to travel an extra angular distance ($\delta\phi_{prec}$: precession angle) in between two successive perihelion points. With the weak field approximation \cite{Parth}, in Schwarzschild case, the precession angle $\delta\phi_{prec}$ can be written as,    $$\delta\phi_{prec}=\frac{6\pi M_{TOT}^2}{h^2}\,\, .$$

As we previously mentioned, in this paper we compare the bound orbits in JNW and Schwarzschild spacetimes, and the bound orbits in a spacetime structure where it is internally BST and externally Schwarzschild spacetime. We basically show how a freely falling particle, with a particular angular momentum and total energy, moves in Schwarzschild spacetime, JNW spacetime and in the BST spacetime structure. In Fig.~(\ref{precgen}), it is shown that the minimum value of effective potentials in Schwarzschild and BST spacetimes are inside the matching radius $R_b=1$. It is possible to have particles' whole trajectory inside the matching radius $R_b$ when the following inequality holds,
\begin{equation}
    \frac{R_b}{h}>3/2\,\, .
\end{equation}
One can obtain the above inequality by considering the minimum value of effective potential of BST spacetime inside the matching radius $R_b$. With the above inequality, a particle needs to have certain amount of total energy to be inside the interior BST spacetime. From the above condition, it can be understood that $\frac{R_b}{h}$ can have arbitrary large values which gives bound trajectories of particles inside the BST spacetime. However, to compare with bound orbits in Schwarzschild spacetime, we need angular momentum per unit rest mass of the particle $h>\sqrt{3}M_0R_b$, which is explained in \citep{Parth}.

The approximate solutions of the orbit eq.~(\ref{orbiteqBST}), corresponding to the BST spacetime can also be written for small values of eccentricity $e$. We can write the eq.~(\ref{orbiteqBST}) in the following form,
\begin{equation}\label{approxBST}
\tilde{u}\frac{d^2\tilde{u}}{d\phi^2}+(1-M_0)\tilde{u}^2=C_{\delta}\tilde{u}^{2\delta}\,\, ,
\label{orbiteq2}
\end{equation}

where $\delta=\frac12$, $C_{\delta}=\frac{\gamma^2R_b^2}{4h^2}$ and $\tilde{u}=uR_b$.  The solution of eq.~(\ref{approxBST}) can be written as,
\begin{equation}
\tilde{u}=\frac{1}{p}\left[1+e\cos(m\phi)+O(e^2)\right]\,\, .
\label{approxorbit}
\end{equation}
Using eq.~(\ref{orbiteqBST}) and eq.~(\ref{approxorbit}), we can get the following expressions of $p$ and $m$,
\begin{eqnarray}
p&=& \frac{4h^2\beta^2}{\gamma^2 R_b^2}\,\, ,\nonumber\\
m&=&\beta\,\, .
\label{m}
\end{eqnarray}
From the above expression of $m$, it can be seen that for $0< M_0< 1$, $m$ is less than one. Therefore, in BST spacetime, a freely falling massive particle always travels greater than $2\pi$ angular distance in between two successive perihelion points. Therefore, we get Schwarzschild like precession in BST spacetime. Similarly, one can write the approximate solution of orbit eq.~(\ref{orbiteqJNW}) corresponding to JNW spacetime as,
\begin{equation}\label{approxJNW}
\tilde{u}\frac{d^2\tilde{u}}{d\phi^2} + Q\tilde{u} = R\tilde{u}^2 + S\tilde{u}^3\,\, ,
\end{equation}
where, $$Q = \left[\frac{b^2\gamma^2(1-n)}{h^2} - \frac{b^2(2-n)}{2h^2}\right]\,\, ,$$
$$R = \left[\frac{b^2\gamma^2(1-n)(1-2n)}{h^2} - \frac{b^2(2-n)(1-n)}{2h^2} - 1\right]\,\, ,$$
$$S = \left[\frac{3}{2} - \frac{b^2(2-n)(1-n)n}{4h^2} + \frac{b^2\gamma^2(1-2n)(1-n)n}{h^2}\right]\,\, .$$
The solution of eq.~(\ref{approxJNW}) can be written as,
\begin{equation}
\tilde{u}=\frac{1}{p}\left[1+e\cos(m\phi)+O(e^2)\right]\,\, .
\label{solapprox}
\end{equation}
where, p and m are positive number. Using eq.~(\ref{orbiteqJNW}) and (\ref{approxJNW}) we can write the expression of p and m by neglecting higher order terms of $e$,
\begin{eqnarray}
   p_{\pm}=\frac{R\pm\sqrt{R^2+4QS}}{2Q}\,\, ,\label{pJNW}\\
   m=\sqrt{Qp-2R-\frac{3S}{p}}\label{mJNW}\,\, ,
\end{eqnarray}
Using eq.~(\ref{mJNW}), one can show that in JNW spacetime, we have two types of precession. A freely falling particle, in this spacetime, can travel greater or less than $2\pi$ angular distance in between two successive perihelion points.

\section{Results and Discussion}
\label{Result}
From the approximate solution of the orbit equation of BST, we get $m=\beta$. As $\beta<1$ in BST spacetimes, a freely falling massive particle have only Schwarzschild like perihelion precession. In fig.~(\ref{orbitBSTM05}, \ref{orbitBSTM01}), the Schwarzschild like precession of the particle orbit in BST is shown, where for fig.~(\ref{orbitBSTM05}) we take $M_0=0.05, h=0.1, E=-0.03$ and for fig.~(\ref{orbitBSTM01}) we take $M_0=0.1, h=0.3, E=-0.01$. In fig.~(\ref{H8},\ref{H88}), the shape of particle orbit is shown, where the particle crosses the matching radius $R_b$. It can be seen that due to the effect of external Schwarzschild spacetime, the shape of the bound orbit changes. In fig.~(\ref{H88}), one can see that in Schwarzschild spacetime no bound orbit is possible. For $M_0=0.333, h=0.4, E=-0.044, R_b=1$, we can see that in Schwarzschild spacetime only the plunge orbit is possible. However, due to the presence of internal BST spacetime, a particle can have bound trajectory, although it has no bound orbits in Schwarzschild spacetime. 
 
From the approximation solution of JNW spacetime, we get the expression of $m$ as written in eq.~(\ref{mJNW}). In fig.~(\ref{re1},\ref{re2}), using eq.~(\ref{mJNW}), we show the region of $m>1$ and $m<1$ for the parameter space of scalar charge $q$ and ADM mass $M$. In fig.~(\ref{M0025},\ref{M025}), we show the effective potentials and the orbits of particles in Schwarzschild and JNW spacetimes for different values of scalar field charge ($q$) and ADM mass $M$. In the fig.~(\ref{orbitJNWq01},\ref{orbitJNWq05}), we can see that for $M = 0.025$, and $q = 0.06$, a freely falling massive particle, in JNW spacetime, has Schwarzschild like precession. On the other hand, if we increase the scalar field charge to $q=0.5$, orbit starts precessing in reverse direction of particle motion. 

One can verify that the points $M=0.025, q=0.06$ and $M=0.025, q=0.5$ lie inside the $m<1$ and $m>1$ regions in fig.~(\ref{re2}) respectively. In fig.~(\ref{orbitJNWq4},\ref{orbitJNWq10}), it can be seen that for $M = 0.25$, $q = 3$ we get Schwarzschild like precession , and for $M = 0.25$, $q = 10$ we get the reverse precession of the orbits of a particle. In fig.~(\ref{re1}), one can verify that the point $M = 0.25$, $q = 3$ lies inside $m<1$ region and $M = 0.25$, $q = 10$ lies inside $m>1$ region. From $b=2\sqrt{M^2+q^2}$, we can get the corresponding values of $b$ for fig.~(\ref{orbitJNWq01},\ref{orbitJNWq05},\ref{orbitJNWq4},\ref{orbitJNWq10}) as $b=0.206, 1.00125, 6.00021, 20.0001$ respectively. 

We know that at $r=b$, JNW spacetime has a strong curvature naked singularity. Therefore, one can see that in fig.~(\ref{orbitJNWq05}), a freely falling particle in JNW spacetime can have bound orbits very close to the central naked singularity. For the above mentioned values of $b$, we always have $n<0.5$. Therefore, for these cases, stable circular timelike orbits, passing through any point in JNW spacetime, are possible.

\section{Conclusion}\label{result}
In the present paper, we investigated the timelike  trajectories of particles in the Schwarzschild, BST and JNW spacetimes, in order to investigate the causal structure of these spacetimes, and to 
understand the important distinguishable properties between them.
This was mainly to understand the presence or otherwise of the event horizon, and possible physical implications in either case.
Following are some of the important differences which can be identified from the nature of the timelike trajectories in the black hole and naked singularity spacetimes: 

\begin{itemize}
    \item In the Schwarzschild spacetime, the perihelion precession of the bound timelike trajectories of a particle is always in the direction of the particle motion. As we know, for BST, we have $m=\beta$ and $\beta<1$. Therefore, like the Schwarzschild spacetime, in BST, particle orbits precess in the direction of particle motion. However, for JNW naked singularity spacetimes, $m$ can be greater or less than one, which is shown in fig.~(\ref{re1},\ref{re2}). Therefore, unlike Schwarzschild spacetime, in JNW spacetime, the particle bound trajectories can precess opposite to the direction of particle motion.

    \item In the Schwarzschild spacetime, there exists an Innermost Stable Circular orbit (ISCO) at $r=6M_{TOT}$, below which no stable circular orbits are possible. On the other hand, in JNW spacetime, for $0.5<n<1$, there exists an ISCO. However, for $0.447<n<0.5$, other than $r_{-}<r<r_{+}$ (eq.~(\ref{isco1},\ref{osco})), stable circular orbits of any radius are possible. For $0<n<0.447$, there are no constrains on the radius of the stable circular orbits of particles in JNW spacetime. We also show that stable circular orbits of any radius are possible in BST. In \cite{Parth}, it is shown that the Joshi-Malafarina-Narayan (JMN) spacetimes which have naked singularity at the center, also have no ISCO. These differences can cause distinguishable properties of accretion disk, which could be possibly detectable by our modern detectors available today \cite{Joshi:2013dva},\cite{Shaikh:2018lcc}.
        
\item In fig.~(\ref{VeffJNWq01},\ref{VeffJNWq05},\ref{VeffJNWq10},\ref{VeffJNWq4}) and in fig.~(\ref{veffBSTM01},\ref{veffBSTM05}), we show that the effective potentials of JNW and BST spacetimes become positive infinite at the center. Therefore, in these two spacetimes, a freely falling massive particle with non-zero angular momentum cannot reach the center and with a certain amount of total energy a particle can be scattered by the infinite potential barrier. However, unlike BST and JNW cases, the effective potential in the Schwarzschild spacetime becomes negative infinity at the center. Therefore, a massive particle with non-zero angular momentum and suitable total energy can plunge into the center.
\end{itemize}

For a better understanding of causal structure and the mass and dynamics of the galactic center, we need to study the timelike and lightlike geodesic behaviour around the galactic center. As we know, GRAVITY and SINFONI are continuously eyeing up the Milkyway galactic center to get important observational data of stellar motion around the center. There are many papers where the bound timelike orbits of particles around black holes and naked singularities are investigated \cite{levin1,Glampedakis:2002ya, Chu:2017gvt, Dokuchaev:2015zxe,Borka:2012tj,Martinez:2019nor,Fujita:2009bp,Wang:2019rvq,Suzuki:1997by,Zhang:2018eau,Pugliese:2013zma,Farina:1993xw,Dasgupta:2012zf,Shoom:2015slu,Eva, Eva1, Eva2, tsirulev}. In such a  context, we show in this paper that the timelike geodesics of a freely falling particle in the JNW and BST naked singularity spacetimes can be significantly different from the timelike geodesics in Schwarzschild spacetime.


\end{document}